\documentclass[epj]{svjour}
\usepackage[utf8]{inputenc}
\usepackage{graphicx,colordvi}
\usepackage{amssymb}
\usepackage{xcolor}
\everymath{\displaystyle}   
\begin{document}
\title{\bf On the Stability of Superheavy Nuclei}
\author{Krzysztof Pomorski\inst{1}\thanks{\emph{Email address:}
 Krzysztof.Pomorski@umcs.pl} \and Artur Dobrowolski\inst{1} \and 
Bo\.zena Nerlo-Pomorska\inst{1} \and Micha{\l} Warda\inst{1} \and 
Johann Bartel\inst{2} \and Zhigang Xiao\inst{3} \and Yongjing Chen\inst{4}
\and Lile Liu\inst{4} \and Jun-Long Tian\inst{5} \and Xinyue Diao\inst{3}
}                     
%
%
\institute{Institute of Physics, Maria Curie Sk{\l}odowska University, 20-031 
Lublin, Poland \and Institut Pluridisciplinaire Hubert Curien, CNRS-IN2P3, 
67-200 Strasbourg, France \and Department of Physics, Tsinghua University, 
100084 Beijing, China \and China Institute of Atomic Energy, 102413 Beijing, 
China \and School of Physics and Electrical Engineering, Anyang Normal
University, 455000 Anyang, China }
\date{Received: date / Revised version: date}
%
\abstract{Potential energy surfaces of even-even superheavy nuclei are
evaluated within the macroscopic-microscopic approximation. A very
rapidly converging analytical Fourier-type shape parametrization is used to
describe nuclear shapes throughout the periodic table, including those of
fissioning nuclei. The Lublin Strasbourg Drop and another effective
liquid-drop type mass formula are used to determine the macroscopic part of 
nuclear energy. The Yukawa-folded single-particle potential, the
Strutinsky shell-correction method, and the BCS approximation for including
pairing correlations are used to obtain microscopic energy corrections.
The evaluated nuclear binding energies, fission-barrier heights, and $Q_\alpha$
energies show a relatively good agreement with the experimental
data. A simple one-dimensional WKB model \`a la \'Swi{\c a}tecki is used to
estimate spontaneous fission lifetimes, while $\alpha$ decay probabilities are
obtained within a Gamow-type model.
} 
%
\maketitle
%
\section{Introduction}

Superheavy nuclei (SHN) have been a challenge for nuclear physicists, both
theoreticians and experimentalists, for the last five or six decades, but
speculations about their existence go back to the end of the 19th and the
beginning of the 20th century \cite{xx1,xx2}. An extensive review of the
properties of these nuclei, including the papers dealing with this subject can
be found in Ref.~\cite{SPo07} and many other review papers as e.g.\
\cite{rev1,rev2,rev3}, which allow us to avoid giving a long list of theoretical
and experimental papers related to SHN. We shall instead concentrate on the
problem of extrapolating what we know from lighter nuclei to the region of
superheavy elements. We would like, in particular, to demonstrate that a
reliable prediction of nuclear ground-state masses is crucial for a dependable
description of spontaneous fission and $\alpha$-decay probabilities. All
calculations reported in this paper are based on the 
macroscopic-microscopic(mac-mic) model \cite{NTS69}, and we show how using 
different modern liquid-drop (LD) type models impacts on
our predictions of masses and fission barrier heights
of the SHN. An analysis of the single-particle spectra obtained in different
mean-field and self-consistent calculations shows that the magic numbers
predicted for the region of SHN are very contingent on the model used. To take
into account the degeneracy of single-particle levels turns out to be essential,
when analyzing nuclear spectra. We will show that the nuclear
shell-correction energy is, in this respect, a much better-suited tool to find
the magic numbers in a given region of nuclei. 

Our paper is organized as follows. The Fourier
shape parametrization \cite{PNB15} that has been shown \cite{SPN17,PNB18} 
to provide an excellent description of the shape of nuclei
throughout the nuclear chart, including the very elongated and necked-in forms 
that appear in very heavy fissile nuclei, is presented in some detail in section
2. Section 3 introduces two models, namely the well known Lublin-Strasbourg Drop
(LSD) \cite{PDu03} and one of the drop models presented in 2006 by Moretto et
al. \cite{MLP12}, to describe the macroscopic nuclear energy. The potential
energy surfaces (PES) of the SHN are evaluated including the microscopic shell
and pairing energy corrections. The analysis of the PES's allows to determine
the fission barrier heights and $Q_\alpha$ energy values which we present in
section 4. Spontaneous fission lifetimes as obtained in a simple WKB type model 
are compared in section 5 with the available experimental data, while 
$\alpha$-decay lifetimes evaluated in a Gamow-type model are presented in
section 6. Section 7 finally gives some conclusions and outlooks.


\section{Fourier shape parametrization}

A proper low-dimensional description of the shape of a nucleus that can undergo
fission is probably one of the most challenging tasks with which nuclear 
physicists have been confronted since the early paper of Bohr and Wheeler 
\cite{BWh39} on nuclear fission theory. Many different parametrizations have 
been proposed to describe the shapes of deformed nuclei (see Ref.~\cite{HMy88} 
for an extensive review of frequently used parametrizations). In the present 
paper, we are using a straightforward and rapidly convergent Fourier type 
parametrization as first presented in Refs.\cite{PNB15,SPN17}.
\begin{figure}[htb]
\includegraphics[width=0.95\columnwidth]{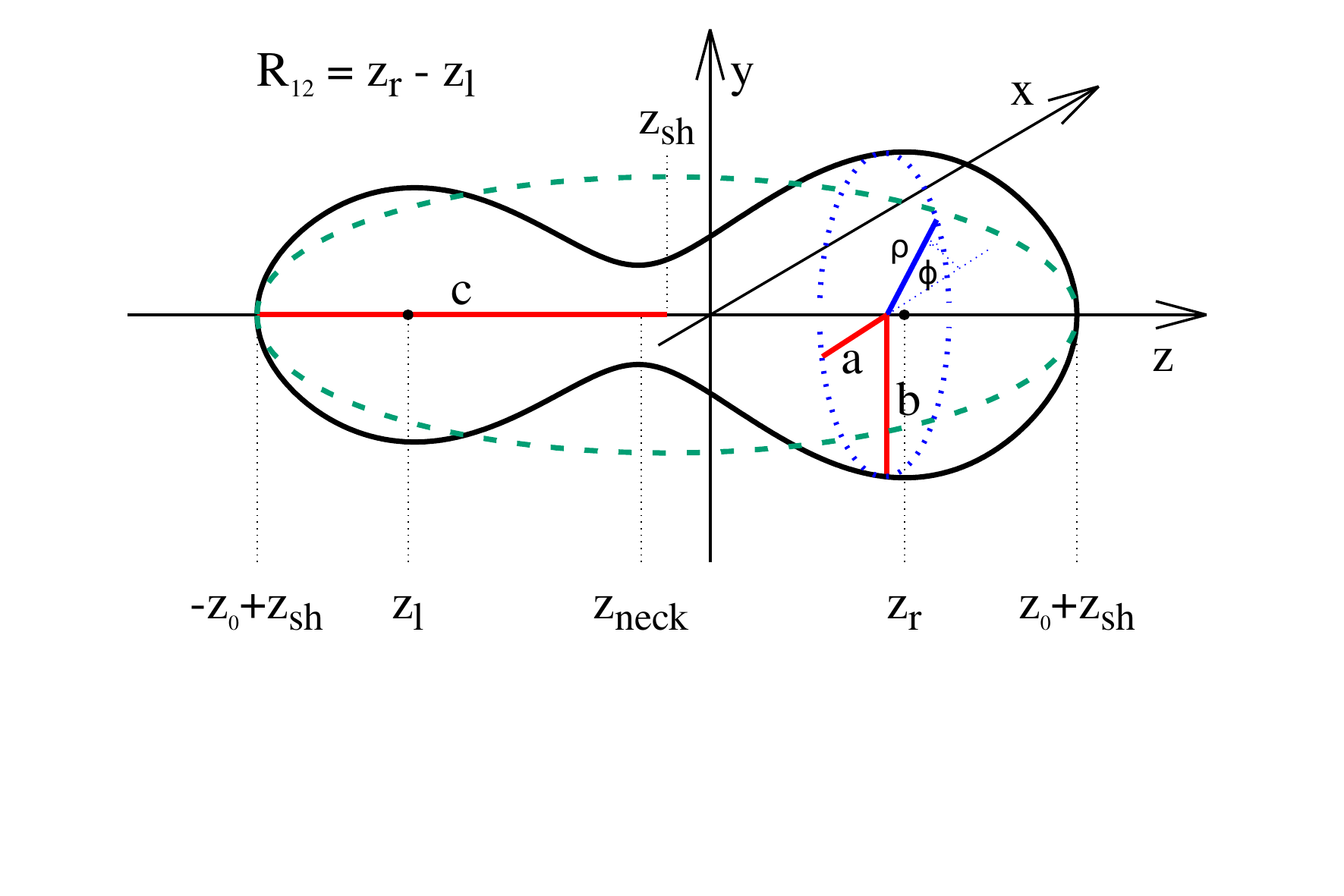}\\[-2ex]
\vspace{-1.5cm}
\caption{Shape of a very elongated fissioning nucleus.}
\label{shape}
\end{figure}

A typical shape of a strongly elongated nucleus is shown in Fig.~\ref{shape},
where the distance $\rho_s(z)$ from the z-axis to its surface is
displayed as a function of the $z$ coordinate, while the dashed line shows a
spheroid having the same length and volume. We assume that for axially symmetric
shapes the function $\rho_s(z)$ is described by the following Fourier series: 
\begin{equation}
\begin{array}{ll}
 \rho_s^2(z)\!=\!R_0^2\! \sum\limits_{n=1}^\infty &\left[
 a_{2n} \cos\left(\frac{(2n-1) \pi}{2} \, \frac{z-z_{sh}}{z_0}\right)\right.\\
 &+\left. a_{2n+1} \sin\left(\frac{2 n \pi}{2} \, \frac{z-z_{sh}}{z_0}\right)
 \right]~,
\end{array}
\label{rhos}
\end{equation} 
where the expansion coefficients $a_\nu$ are treated as free deformation
parameters. The half-length $z_0$ of the shape is evaluated from the
$a_\nu$ parameters imposing the condition that the volume of the nuclear shape
needs to be conserved, while $z_{\rm sh}$ ensures that the center of mass of 
the deformed nucleus is located at the origin of the coordinate system, with 
$R_0$ being the radius of the spherical nucleus having the same volume.

Non-axial shapes can easily be taken into account by assuming that the
cross-section of the nuclear surface perpendicular to the $z$-axis has the form
of an ellipse with half-axis $a$ and $b$. One can then define a non-axiality
deformation parameter $\eta$ as:
\begin{equation}
   \eta=\frac{b-a}{a+b}~.
\label{eta}
\end{equation}
In the case of a non-axial nuclear shape the volume conservation condition
implies that $\rho_s^2(z)=a(z)b(z)$. The following equation then gives the 
non-axial shape:
\begin{equation}
   \varrho^2(z,\varphi) = \rho^2_s(z) \, 
   \frac{1-\eta^2}{1+\eta^2+2\eta\cos(2\varphi)}~,
\label{nrho}
\end{equation}
where one assumes that the non-axiality parameter $\eta$ is independent on $z$.
The rapid convergence of the Fourier parametrization (\ref{rhos}) for any 
realistic nuclear shape is striking, as has been demonstrated in Ref.~
\cite{SPN17}.

In practical calculations, instead of the $a_\nu$ deformation parameters, 
it is more convenient to use the following combinations:
\begin{equation}
\begin{array}{l}
q_2 =a_2^{(0)}/a_2 - a_2/a_2^{(0)} \,\,,\\
q_3=a_3 \,\,,\\
q_4=a_4+\sqrt{(q_2/9)^2+(a_4^{(0)})^2} \,\,,\\
q_5=a_5-(q_2-2)a_3/10 \,\,, \\
q_6=a_6-\sqrt{(q_2/100)^2+(a_6^{(0)})^2} \,\,,
\end{array}
\label{qi}
\end{equation}
where the $a_{2n}^{(0)}=(-1)^{n-1}32/[\pi(2n-1)]^3$ are the Fourier expansion 
coefficients of a sphere.
The new $q_\nu$ deformation parameters have the following physical interpretation:\\
$\star\;$ $q_2$ : elongation of the nucleus,\\
$\star\;$ $q_3$ : left-right asymmetry,\\
$\star\;$ $q_4$ : neck formation,\\
$\star\;$ $q_5$ and $q_6$ regulate the deformation of fission fragments.
                                                                    \\[ 1.0ex]
These parameters were chosen in such a way that the liquid-drop path to fission 
corresponds approximately to $q_3=q_4=q_5=q_6=0$ and the spherical shape is 
obtained when all $q_\nu$ vanish.


\section{Total nuclear energy}

We have performed calculations using the mac-mic
model approach of the total nuclear energy \cite{MSw66}:
\begin{equation}
E_{\rm tot} = E_{\rm mac}+ E_{\rm shell}^{\rm p}
+ E_{\rm shell}^{\rm n} + E_{\rm pair}^{\rm p} + E_{\rm pair}^{\rm n}~,
\label{etot}
\end{equation}
where $E_{\rm mac}$ is the macroscopic part of the energy, while $E_{\rm shell}$
and $E_{\rm pair}$ describe the microscopic shell and pairing energy corrections
for protons and neutrons.


\subsection{Liquid-drop type mass formula}

In what follows, two types of macroscopic models have been used: the 
Lublin-Strasbourg Drop (LSD) \cite{PDu03} which has been shown to represent one 
of the most performant liquid-drop type models on the one hand, and one of
the five LD formulas proposed in 2012 by Moretto et al. \cite{MLP12}, namely 
the version (i) which has, contrary to most existing LD formulas, the same
isospin-square dependence of the volume and surface terms and does not contain
any curvature contribution. Both these modern versions of the nuclear LD model
which have the particularity to reproduce rather well not only nuclear masses but
also the fission-barrier heights in the actinide region \cite{Pom13}, will allow
us to compare their extrapolations from the known nuclear region to one of the
superheavy nuclei.

The LSD model \cite{PDu03} contains, in addition to the traditional volume, 
surface, and Coulomb terms, a curvature and a congruence (Wigner) energy term
\begin{equation}
\begin{array}{rl}
  M_{\rm LSD}&\hspace{-1mm}(Z,N;q_i) = Z M_{\rm H} + N M_{\rm n}
   - b_{\rm elec} \, Z^{2.39}                              \hfill\\[1.2ex]
      &+\, b_{\rm vol}\;\,(1 - \kappa_{\rm vol} \; I^2\,)\,A     \\[1.2ex]
      &{ +\, b_{\rm surf}\,(1 - \kappa_{\rm surf} I^2\,)\,A^{2/3}
         B_{\rm surf}(q_i) }                               \\[1.2ex]
      & +\, b_{\rm cur}\;\,(1 - \kappa_{\rm cur} \; I^2\,)\,A^{1/3}
         B_{\rm cur}(q_i)                                 \\[1.4ex]
      &{ +\, \frac{3}{5} \,  \frac{e^2Z^2}{r_0^{\rm ch} A^{1/3}}\,
         B_{\rm Coul}(q_i)} -\, C_{4}\,\frac{Z^2}{A}       \\[2ex]
      & + E_{\rm cong}(Z,N) + E_{\rm odd} + E_{\rm micr}(Z,N;q_i)~.   
\end{array}
\label{lsd}
\end{equation}
Here the nuclear deformation is identified by the parameter set $\{q_i\}$ in
(\ref{qi}), $I=(N-Z)/A\;$ is the reduced isospin, and $M_{\rm n}$ and $M_{\rm
H}$ are respectively the masses of the neutron and the hydrogen atom. The
coefficient of the electron-shell binding energy is $b_{\rm
elec}=0.00001433\;$MeV, and the odd-even term $E_{\rm odd}$ and the congruence
(Wigner) energy, given as $E_{\rm cong}(Z,N)=10\exp(-4.2\,|I|)$ MeV, are taken
from Ref.~\cite{MSw96}. The ground-state microscopic-energy corrections $E_{\rm
micr}(Z,N;q_i)$ are finally taken from the tables presented by M\"oller at al.
in Ref.\ \cite{MNM95}. The remaining LD parameters of the LSD mass formula have
been adjusted to obtain the best possible fit of the 2766 nuclear masses with
Z$\geq$8 and N$\geq$8, experimentally known by the time of the mass fit of Ref.\
\cite{PDu03}.  

The LD type formulae \cite{MLP12} of Moretto and coworkers have the following
form:
\begin{equation}
\begin{array}{rl}
M_{\rm MLD}&\hspace{-1mm}(Z,N;q_i) = Z M_{\rm H} + N M_{\rm n}\\[1ex]
&\hspace{-9mm}+\left[b_{\rm vol}A+a_{\rm surf}A^{2/3}B_{\rm surf}(q_i)
  +a_{\rm cur}A^{1/3}B_{\rm cur}(q_i)\right]\\[1.5ex]
&\hspace{+15mm}\cdot\left[1-\kappa |N-Z|(|N-Z|+2)/A^2\right] \\[1ex]
&\hspace{-9mm}+a_{\rm Coul}\frac{Z(Z-1)}{A^{1/3}}B_{\rm Coul}(q_i) 
  \pm \frac{\delta}{\sqrt{A}}+E_{\rm micr}(Z,N;q_i).
\end{array}
\label{mld}
\end{equation}
Here the volume, surface, and curvature terms carry the same reduced isospin
dependence, and the electron binding energy is absent. The last two terms
describe the odd-even mass difference and the microscopic correction energy,
which is the same as in Eq.~(\ref{lsd}). The term $|N-Z|(|N-Z|+2)$ origins from
the requirement that the nuclear part of the total energy has to depend on the
square of the isospin of a nucleus: $<\hat\tau^2>=\tau(\tau+1)$, where
$\tau={|N-Z|/2}$, and $<\hat\tau^2>=|N-Z|(|N-Z|+2)/4$. Note that the linear 
term $|N-Z|$ in Eq.~(\ref{mld}) corresponds to the Wigner (congruence) energy
present in the Eq.~(\ref{lsd}). A standard deformation dependence of surface,
curvature and Coulomb energies, absent in the original paper \cite{MLP12}, is
added in the present investigation.

Five different sets of parameters were fitted in Ref.~\cite{MLP12} to 2076
nuclear masses but only one of them $(i)$ with the parameters shown in the table below:
\begin{center}
\begin{tabular}{|c|c|c|c|c|c|}
\hline
             &              &             &        &              &   
                                                                    \\[ -1.5ex]
$a_{\rm vol}$&$a_{\rm surf}$&$a_{\rm cur}$&$\kappa$&$a_{\rm Coul}$&$\delta$
                                                                    \\[  0.9ex]
   [MeV]     &   [MeV]      &   [MeV]     &   --   &      [MeV]   & [MeV]\\ 
\hline
             &              &             &        &              &
                                                                 \\[ -1.6ex]
  -15.597    &   17.32      &    0.0      & 1.8048 &    0.7060    & 11.4
                                                                    \\[  0.9ex]

\hline
\end{tabular}
\end{center}
is able to reproduce the experimental barrier heights with a quality 
similar to the one of the LSD model, as it was shown in Ref.~\cite{Pom13}. 
In what follows, we will only consider this parameter set, labeled MLD 
hereafter, to compare its results with the ones of the LSD mass formula 
and to predict properties of SHN.
\begin{figure}[t!]
\includegraphics[width=1.05\columnwidth]{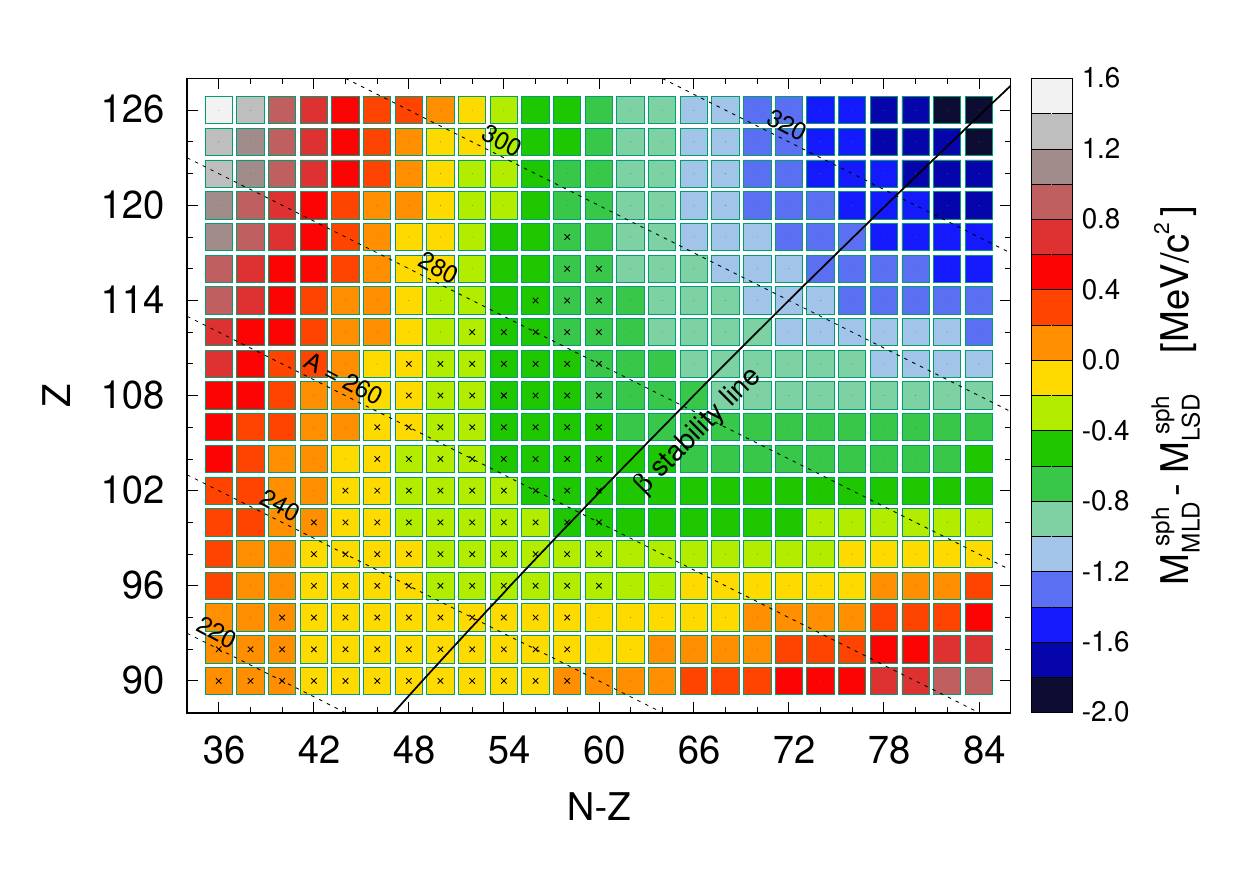}
\caption{Difference between the MLD and LSD estimates of nuclear masses. Crosses
mark the experimentally known isotopes. Constant A values are represented by
dashed lines, while the $\beta$-stable nuclei are to be found around the solid
line.}
\label{mdiff}
\end{figure}
\begin{figure}[h!]
\includegraphics[width=1.05\columnwidth]{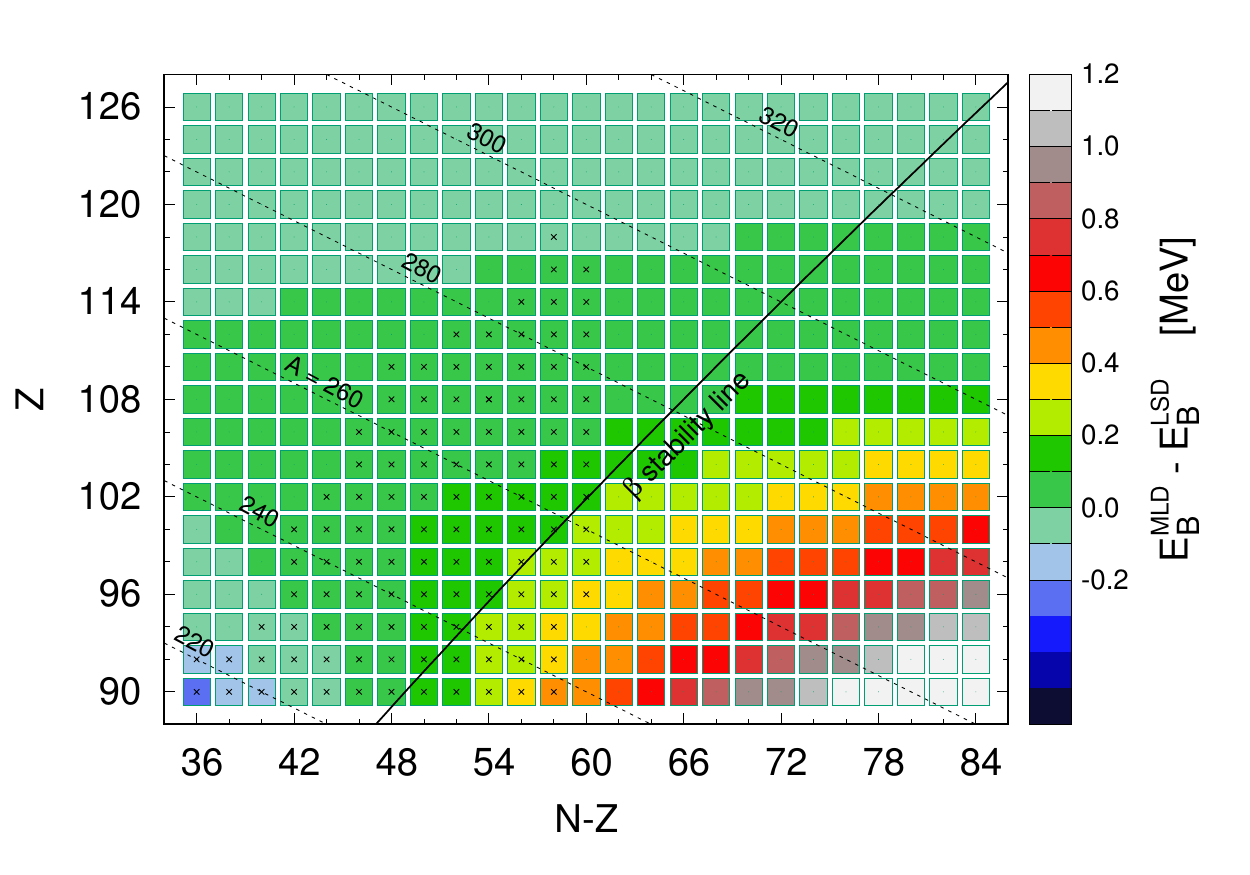}
\caption{Difference between the LD barrier heights evaluated using the
LSD (\ref{lsd}) and MLD (\ref{mld}) formulas.}
\label{bdiff}
\end{figure}
\begin{figure*}[t!]
\includegraphics[width=0.95\columnwidth]{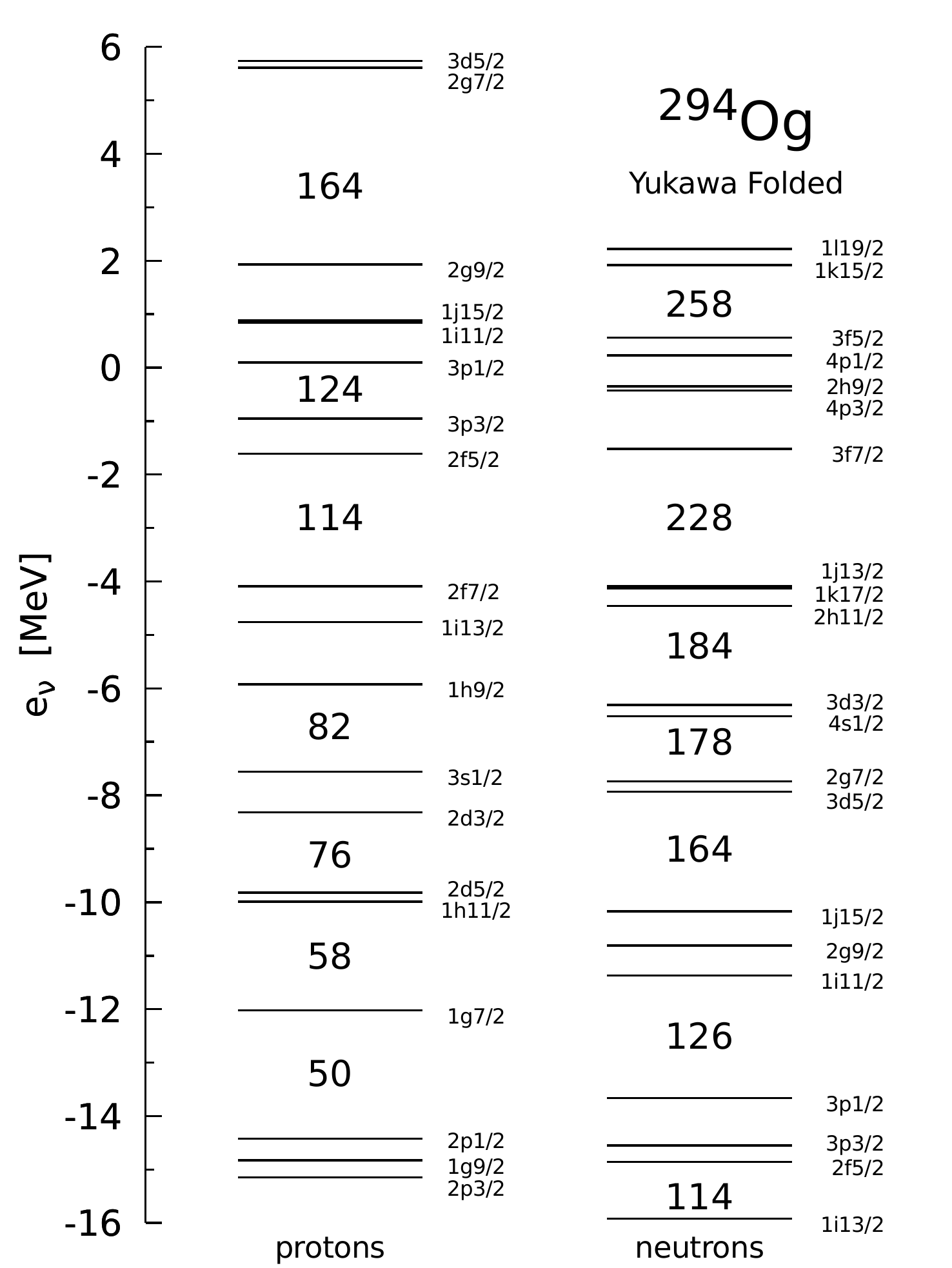}\hfill
\includegraphics[width=0.95\columnwidth]{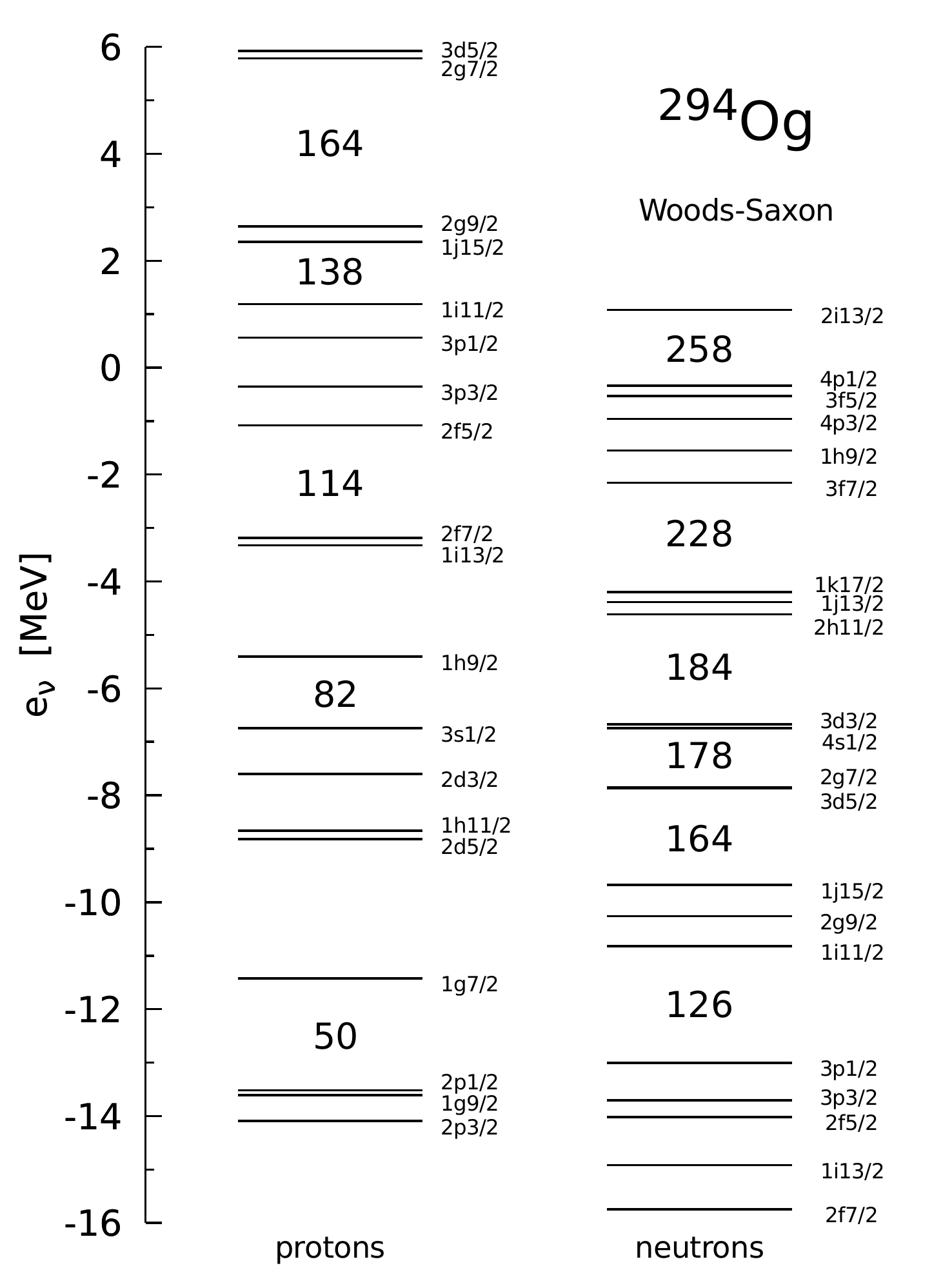}
\caption{Single-particle levels of the spherical nucleus $^{294}$Og evaluated
using the Yukawa-folded (l.h.s.) and the Woods-Saxon (r.h.s) potential.}
\label{splev}
\end{figure*}

The difference between the mass estimates obtained in the MLD and LSD models 
for actinide and superheavy nuclei is shown in Fig.~\ref{mdiff}. Crosses mark
known isotopes. The solid line indicates the $\beta$-stability line, while 
dashed lines correspond to constant A values. The difference between both 
estimates is smaller than 0.2 MeV in actinide nuclei while it becomes slightly 
larger but does not exceed 0.6 MeV around the heaviest known SHN. 
$\alpha$-decay chains correspond to vertical lines in Fig.~\ref{mdiff} with an 
almost constant energy difference between both mass estimates for known elements
with Z$\geq$102, thus indicating that $\alpha$-decay $Q_\alpha$ energies will be
very similar for both LD models in the SHN region.

Liquid-drop fission barrier heights evaluated using the MLD model turn out to be slightly larger than those of LSD, as can be seen in Fig.~\ref{bdiff} but the 
difference does not exceed 0.2 MeV in the superheavy region. 

The very close estimation of the masses and the barrier heights of the SHN
obtained in both modern LD models gives a certain guarantee that our
description of the macroscopic part of the energy is reasonably accurate.


\subsection{Microscopic part of the energy}

Strutinsky type shell corrections are obtained as usual by subtracting the 
average energy $\widetilde E$ from the sum of the energies $e_k$ of the 
occupied single-particle (s.p.) levels
\begin{equation}
  {\rm E_{\rm shell}} = \sum\limits_{\rm occ} e_k - \widetilde E \; .
\label{eq02}
\end{equation}
The energies $e_k$ in (\ref{eq02}) are the eigenvalues of a mean-field
Hamiltonian with a mean-field potential \cite{DNi77}. The
average energy $\widetilde E$ is evaluated using the Strutinsky prescription
\cite{NTS69,Str67}. The pairing-energy correction is determined as the
difference between the BCS energy and the s.p.\ energy sum from
which the average pairing energy is subtracted \cite{NTS69}
\begin{equation}
 E_{\rm pair} = E_{\rm BCS} - \sum\limits_{\rm occ} e_k
            - \widetilde{E}_{\rm pair}\;.
\label{Epair}
\end{equation}
In the BCS approximation, the ground-state energy of a system with an even
number of particles is given by
\begin{equation}
E_{\rm BCS} = \sum_{k>0} 2e_k v_k^2 - G\left(\sum_{k>0}u_kv_k\right)^2 - G\sum_{k>0} v_k^4 -{\cal E}_0^\varphi\;,
\label{EBCS}
\end{equation}
where the sums run over the pairs of s.p.\ levels belonging to the pairing
window defined below. The coefficients $v_k$ and $u_k=\sqrt{1-v_k^2}$ are the
BCS occupation amplitudes, and ${\cal E}_0^\varphi$ is the energy correction due
to the particle number projection performed in the GCM+GOA approximation
\cite{GPo86}
\begin{equation}
{\cal E}_0^\varphi=\frac{\sum\limits_{k>0}[ (e_k-\lambda)(u_k^2-v_k^2)
        +2\Delta u_k v_k +Gv_k^4] / E_k^2}{\sum\limits_{k>0} E_k^{-2}}\;.
\label{Ephi}
\end{equation}
Here, $E_k=\sqrt{(e_k-\lambda)^2+\Delta^2}$ is the quasi-particle energy and
$\Delta$ and $\lambda$ the pairing gap and Fermi energy, respectively. The
average projected pairing energy, for a pairing window, of width $2\Omega$, 
symmetric in energy relative to the Fermi level, is equal to
\begin{equation}
\begin{array}{l}
 \widetilde{E}_{\rm pair}= -\frac{1}{2}\,\tilde{g}\,
 \tilde{\Delta}^2+\frac{1}{2}\tilde{g}\,G\tilde{\Delta}\,
 {\rm arctan}\left(\frac{\Omega}{\tilde\Delta}\right)
  -\log\left(\frac{\Omega}{\tilde\Delta}\right)\tilde{\Delta}\\[3ex]
~~~~~~~~~
+\frac{3}{4}G\frac{\Omega/\tilde{\Delta}}{1+(\Omega/\tilde{\Delta})^2}/
  {\rm arctan}\left(\frac{\Omega}{\tilde{\Delta}}\right)-\frac{1}{4}G ~ ,
\end{array}
\label{Epavr}\end{equation}
where $\tilde{g}$ is the average single-particle level density at the Fermi
surface and $\tilde\Delta$ the average pairing gap corresponding to a pairing
strength $G$ 
\begin{equation}
  \tilde\Delta=2\Omega\exp\left(-\frac{1}{G\tilde{g}}\right)~.
\label{Davr}\end{equation}
All details of the calculation and the parameters used are described in
Ref.~\cite{KDN21}, where an extended set of maps with the potential energy
surfaces (PES) of superheavy nuclei are presented. This pairing energy as 
defined by Eq. (\ref{Epair}) has of course to be evaluated separately for protons
and neutrons as indicated by Eq. (\ref{etot}).
\begin{figure}[h!]
\centerline{\includegraphics[width=\columnwidth]{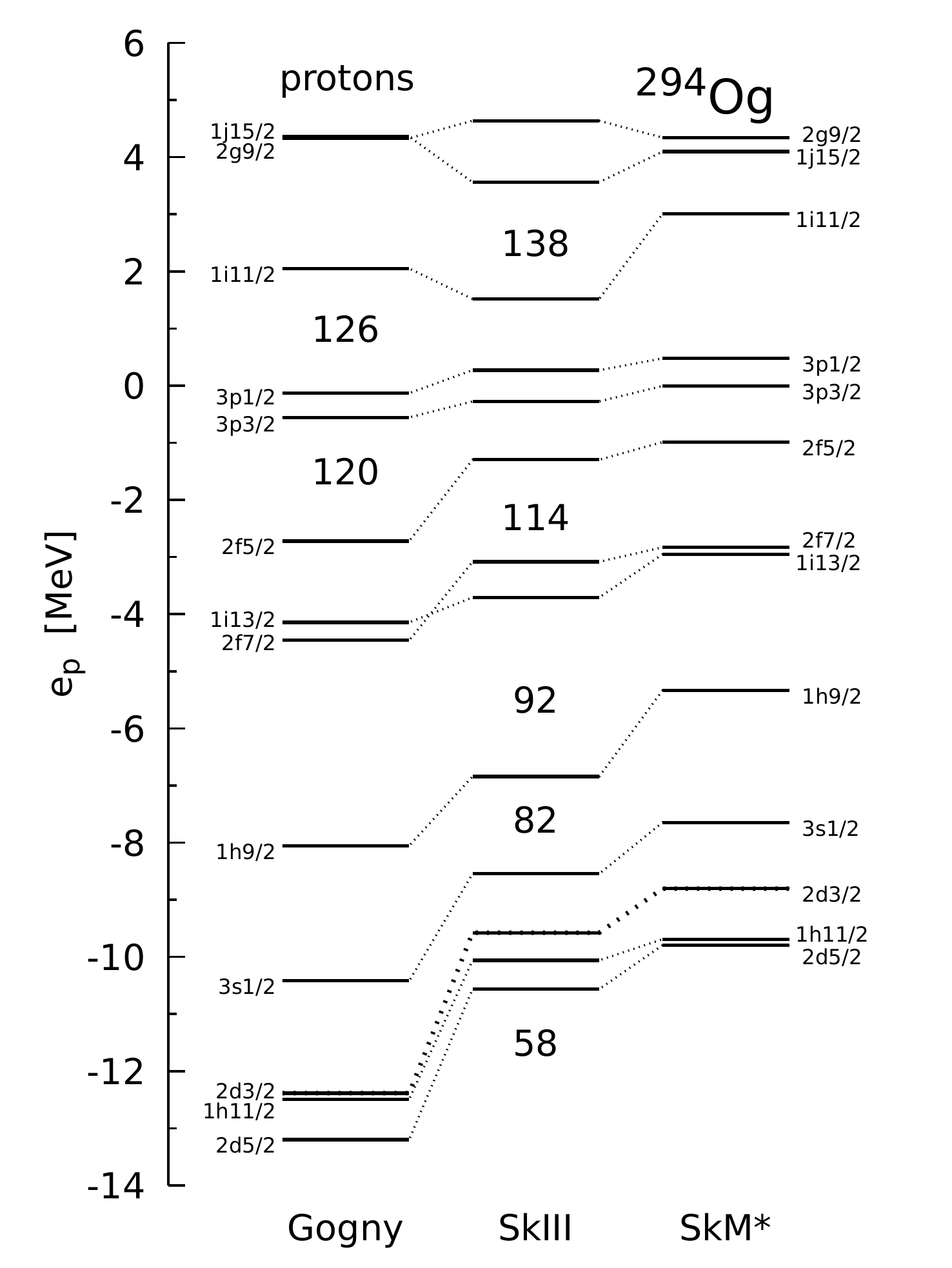}}
\caption{Single-particle proton levels of the spherical nucleus $^{294}$Og
evaluated self-consistently using the SkIII and SkM* Skyrme and Gogny D1S
forces. }
\label{spscp}
\end{figure}
\begin{figure}[h!]
\centerline{\includegraphics[width=\columnwidth]{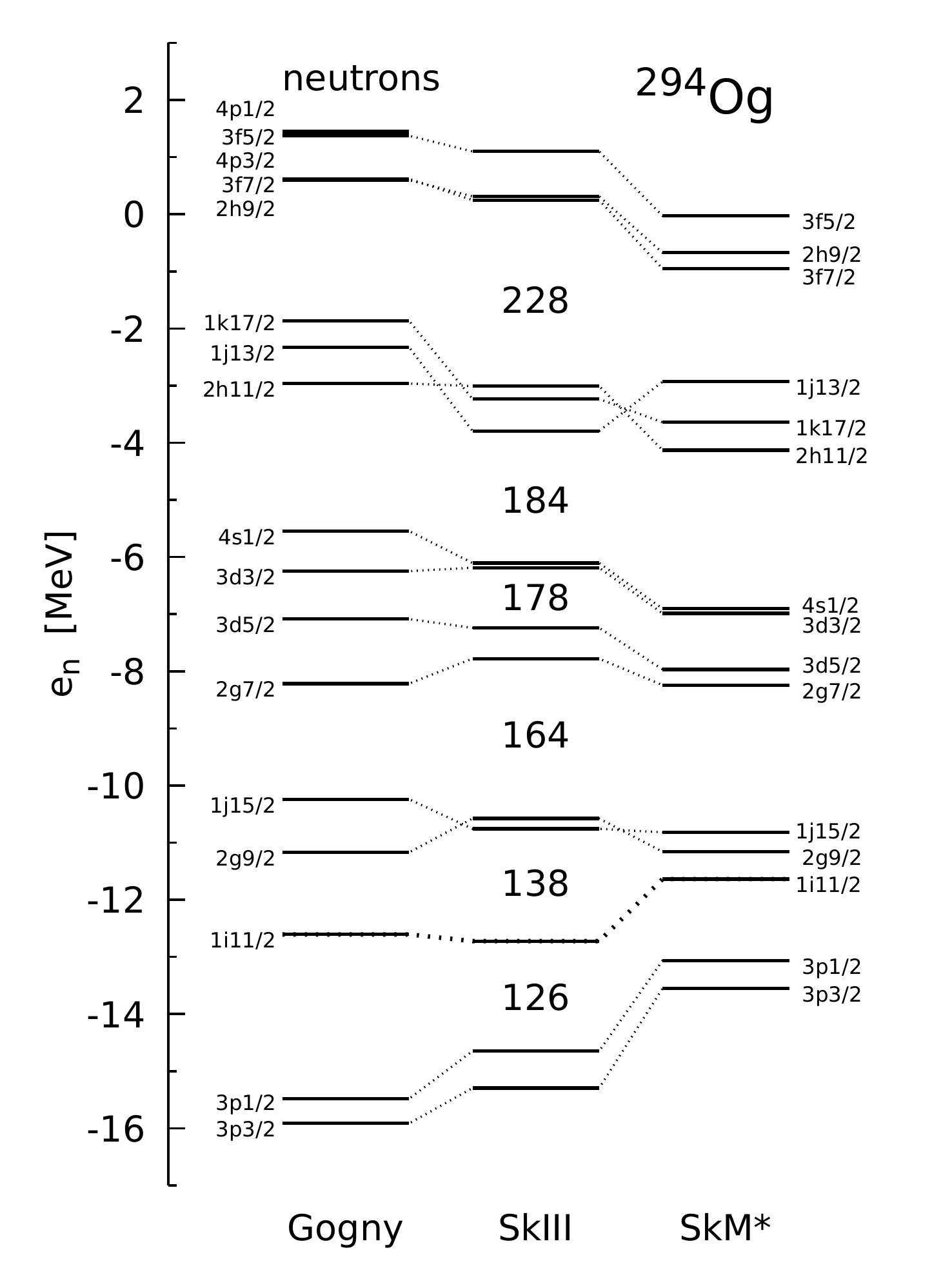}}
\caption{The same as in Fig.~\ref{spscp} but for neutrons. }
\label{spscn}
\end{figure}
\begin{figure*}[t!]
\includegraphics[width=1.05\columnwidth]{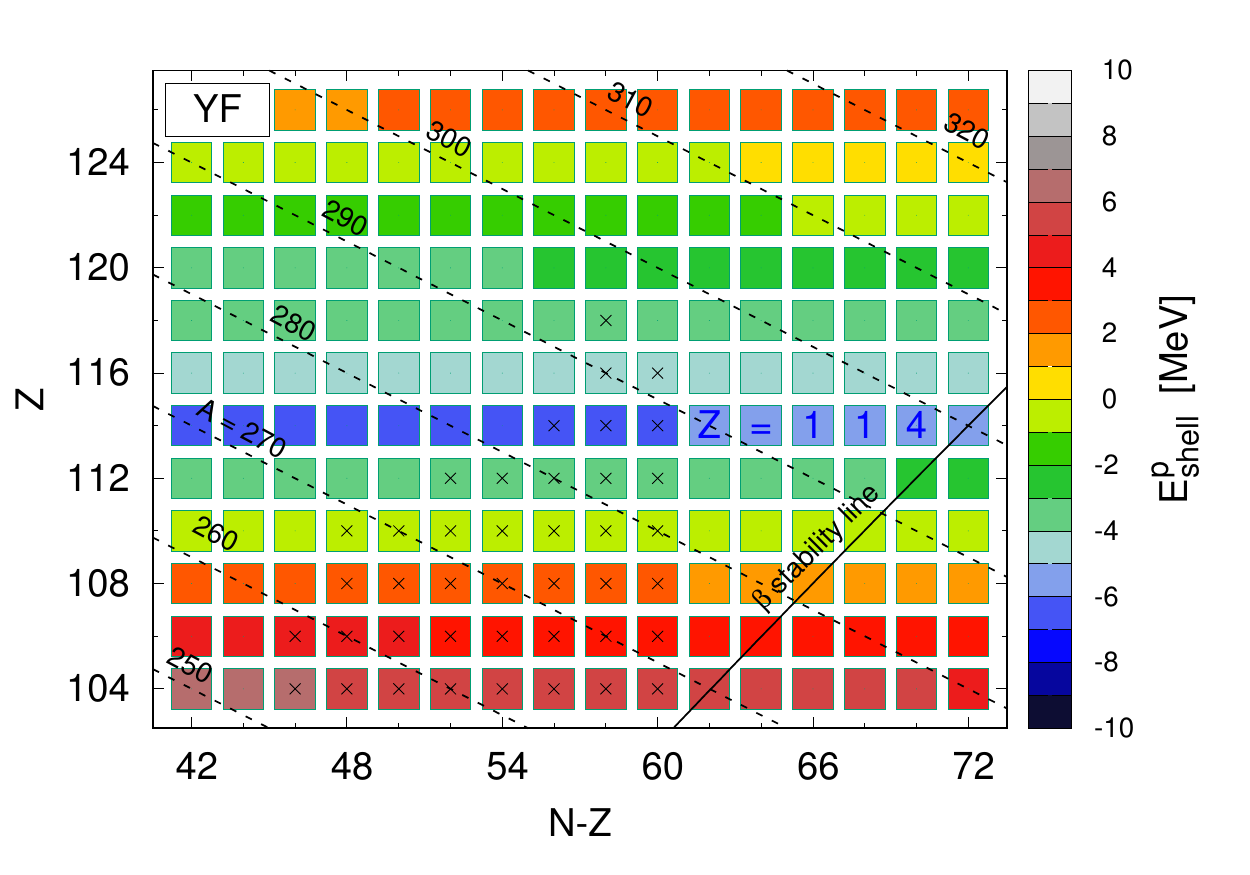}
\includegraphics[width=1.05\columnwidth]{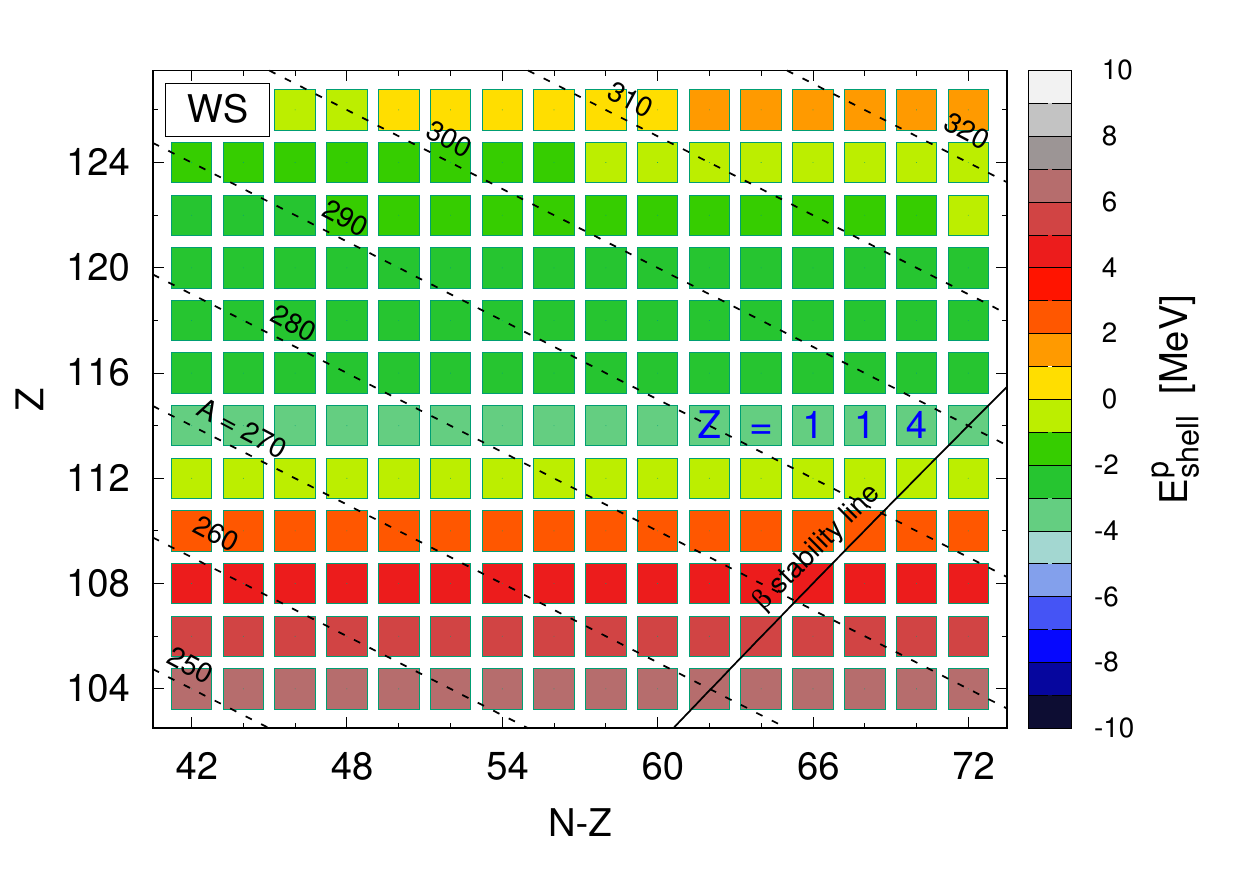}\\
\includegraphics[width=1.05\columnwidth]{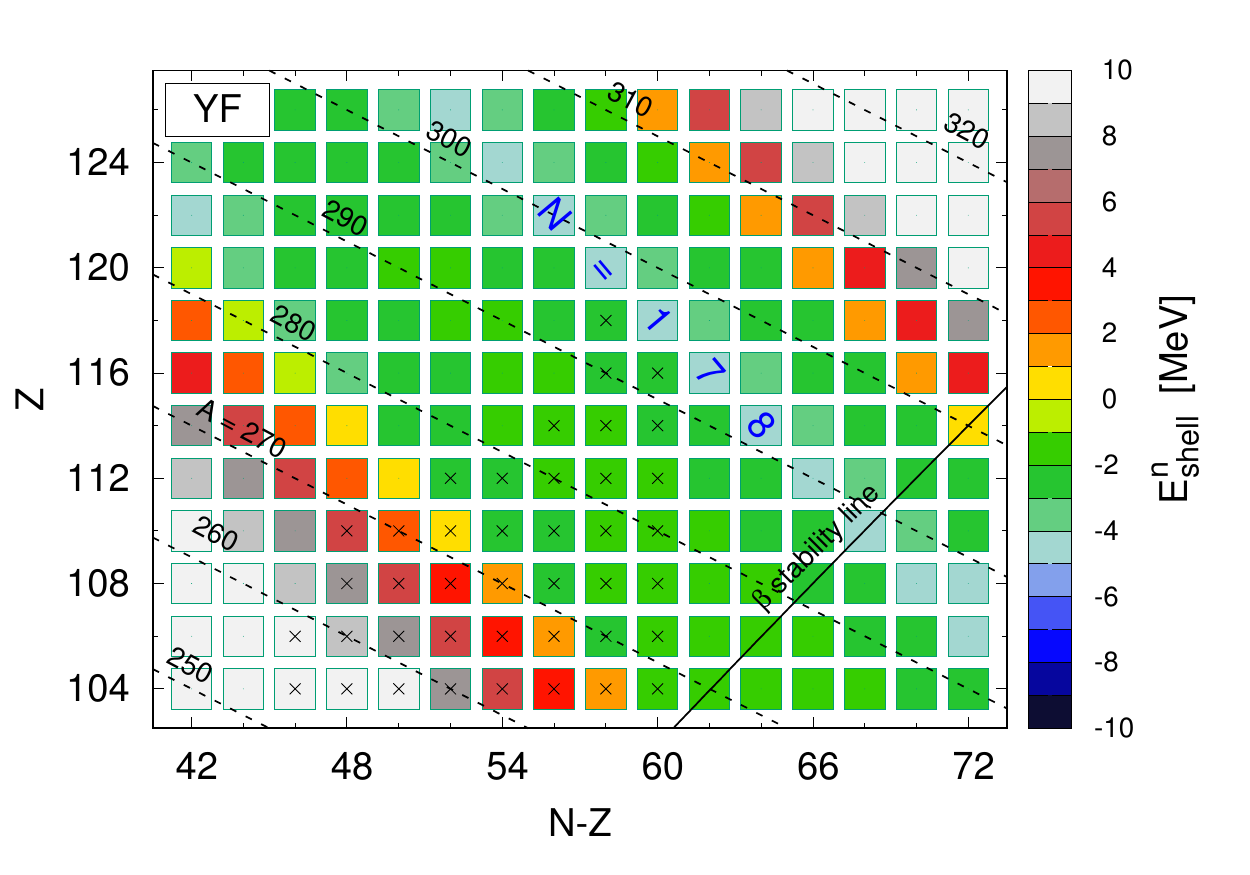}
\includegraphics[width=1.05\columnwidth]{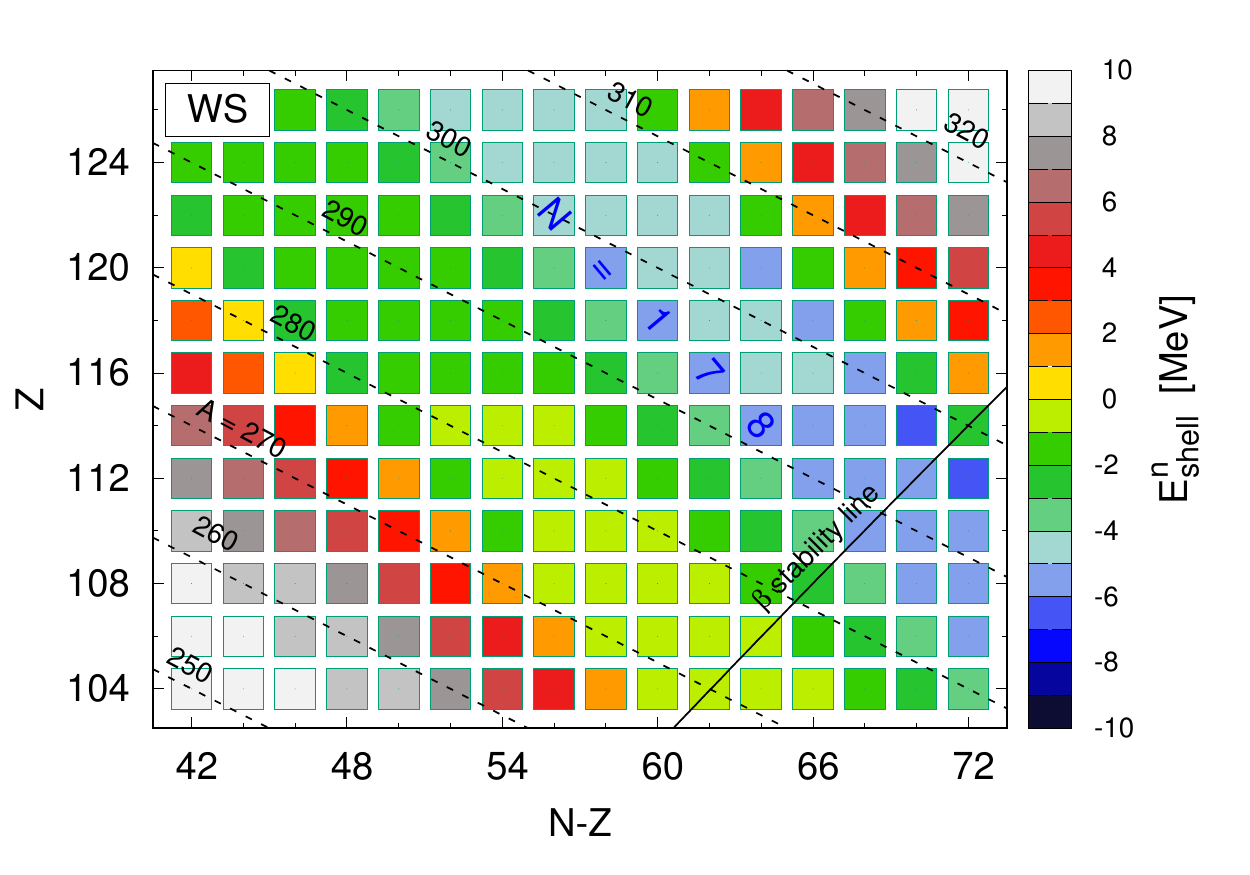}
\caption{Proton (top) and neutron (bottom) shell correction energies of the
spherical actinide and superheavy nuclei obtained using the Yukawa-folded (l.h.s) and the Woods-Saxon (r.h.s) s.p. potentials.}
\label{esh}
\end{figure*}

In our calculation, the single-particle spectra are obtained by diagonalization
of a Hamiltonian with a Yukawa-folded (YF) potential \cite{DNi77,DPB16}
with the same parameters as used in Ref.~\cite{MNM95}. Its s.p.\ proton and
neutron orbitals $|nlj\rangle$ of a spherical $^{294}$Og nucleus are compared
in Fig.~\ref{splev} with the levels evaluated using the Woods-Saxon (WS)
potential with the so-called {\it universal set of parameters} of Ref.\
\cite{CDN87}. Both energy spectra are similar. The main difference is seen in
the proton spectrum above the 2f7/2 levels corresponding to the magic number
Z\,=\,114, where the distance between the orbitals is slightly different.
However, the situation is quite different when one compares these spectra with
the ones evaluated in self-consistent models. In Figs.~\ref{spscp} and
\ref{spscn} they are compared to the proton and neutron spectra obtained
self-consistently using the Skyrme SIII and SkM* and the finite-range Gogny
D1S interaction. As one can see, the order of the proton's orbitals obtained
with the SIII set of parameters is similar to that of the YF potential.
However, the energy distances between the levels are more considerable. The two
other proton s.p.\ spectra obtained with the SkM* and D1S forces show different
orders of the orbitals around the Fermi energy. Also, the energy gap
corresponding to the magic number Z\,=\,114 is smaller in these three 
self-consistent models than in the case of the Yukawa-folded mean-field 
potential. The energy gap corresponding to Z\,=\,126 is most pronounced in
the SkM* spectrum. The situation is similar in the neutron spectra, where energy
gaps are visible around N\,=\,168 and N\,=\,184, but the sequence of orbitals
frequently differs from model to model.

Comparing the different spectra, one should not forget, however, about the
$2j+1$ degeneracy of the orbitals. The contribution of a single orbital to the
shell energy depends obviously on its degeneracy and the energy distance from
the Fermi level. The gaps observed in the energy spectra could be misleading. A
better way to judge the magic numbers would be to compare the shell-correction
energies corresponding to the different numbers of protons and neutrons. The
proton (top) and neutron (bottom) shell correction energies evaluated with the
Yukawa-folded (YF) and the Woods-Saxon (WS) s.p. potentials are presented in
the l.h.s. and r.h.s. columns of Fig.~\ref{esh}, respectively. The proton magic
number at Z\,=\,114 is well visible in the proton shell correction energy which
reaches there around -6 MeV for the YF potential and -3.5 MeV for the SW one. 
Similarly, a neutron magic number N\,=\,178 appears in the neutron shell
correction what is visible in both bottom panels. The neutron shell correction
corresponding to this magic number is around -4 MeV in the YF case and -5.5 MeV
for the WS potential. The shell corrections at the above proton and neutron
magic numbers change slightly with the mass number A. The proton shell energy is
negative for $110\lesssim{\rm Z}\lesssim 124$ and the neutron one for
$164\lesssim {\rm N} \lesssim 184$, which indicates that nuclei in this range of
proton and neutron numbers are spherical in the ground state. Of course, only
the complete macroscopic-microscopic calculation can finally decide whether a
given nucleus is spherical or deformed.

An extended set of 2D cross-sections of the 4D PES's of superheavy nuclei
obtained in our model can be found in Ref.~\cite{KDN21}. The following section
will show only a few examples of such maps.


\section{Barrier heights and $Q_\alpha$ energies}

Calculations have been carried out for the superheavy nuclei with
$104\le {\rm Z}\le 128$ and $250\le {\rm A}\le 324$. The PES have been evaluated
in the 4D deformation parameter space: $(\eta,\,q_2,\,q_3,\,q_4)$. 
\begin{figure}[h!]
\includegraphics[width=0.95\columnwidth]{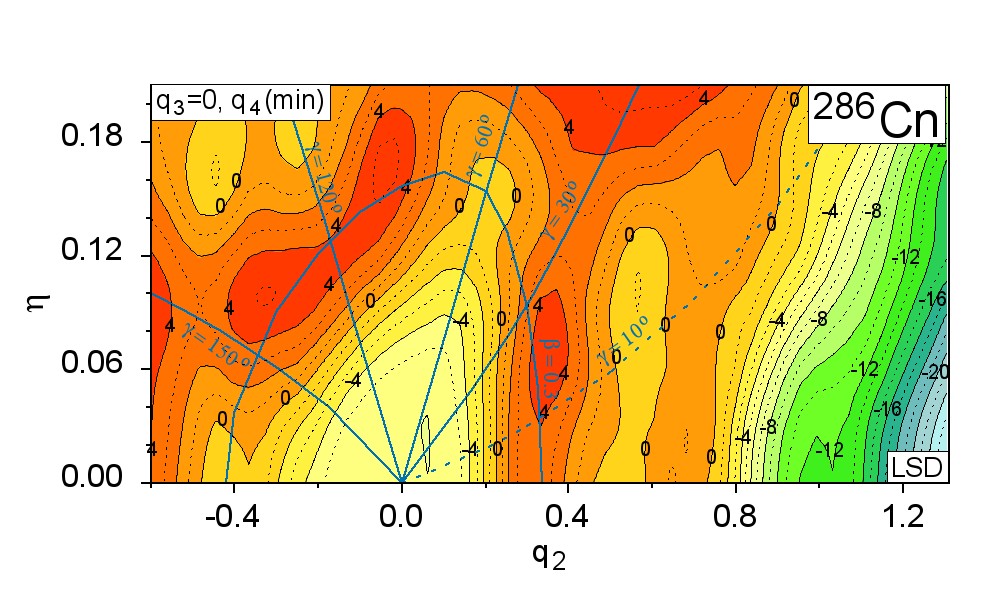}\\
\includegraphics[width=0.95\columnwidth]{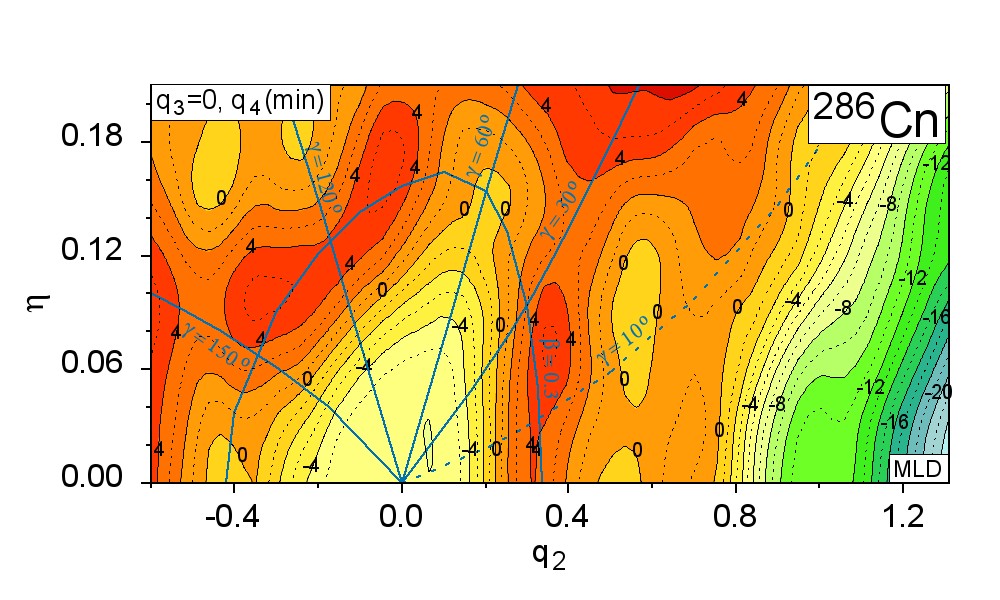}
\caption{PES obtained using the LSD (top) and MLD liquid-drop (bottom) formula
of $^{286}$Cn on the ($q_2,\eta$) plane. }
\label{286Cn21}
\end{figure}

In Fig.\ \ref{286Cn21} the ($q_2,\eta$) cross-section of the 4D potential energy
surface of $\,^{286}$Cn \cite{KDN21} is shown using the LSD (top) and MLD
(bottom) form of LD expression. The energies denoted on the layers are to be
understood as relative to the LD energy of the spherical nucleus. The distance 
between layers (solid lines) is 2 MeV, while the intermediate dashed lines 
correspond to the half-layers. Both cross-sections (for LSD and MLD) in Fig.\
\ref{286Cn21} have been evaluated by imposing left-right symmetry ($q_3=0$) and
minimizing the energy in each ($q_2,\eta$) point with respect to $q_4$. The
green lines marked by $\beta=0.3$ and $\gamma=10,\,30,\,60,\,120,\,150$
correspond to the frequently used ($\beta,\,\gamma$) Bohr deformation parameters
\cite{Boh52}. As one can see, both PES evaluated in different LD models are very
close and give very similar estimates of the saddle point energy. 
The ground state of $^{286}$Cn turns out to be nearly spherical. The fission
valley goes first via oblate ($\gamma=60^o$) shapes, then goes through a
triaxial saddle point at ($\beta \approx 0.38,\,\gamma\approx 30^o$), and a
non-axial second minimum ($\beta \approx 0.55,\,\gamma\approx 15^o)$ to a second
also, non-axial saddle. Then at elongations $q_2\geq 0.8$ the fission valley
returns to axially symmetric shapes ($\eta=\gamma=0$). Such a situation is
typical for all SHN with Z$\geq 110$, which are spherical in the ground-state
(see the collection of PES's in Ref.~\cite{KDN21}).\\
\begin{figure}[h!]
\includegraphics[width=0.95\columnwidth]{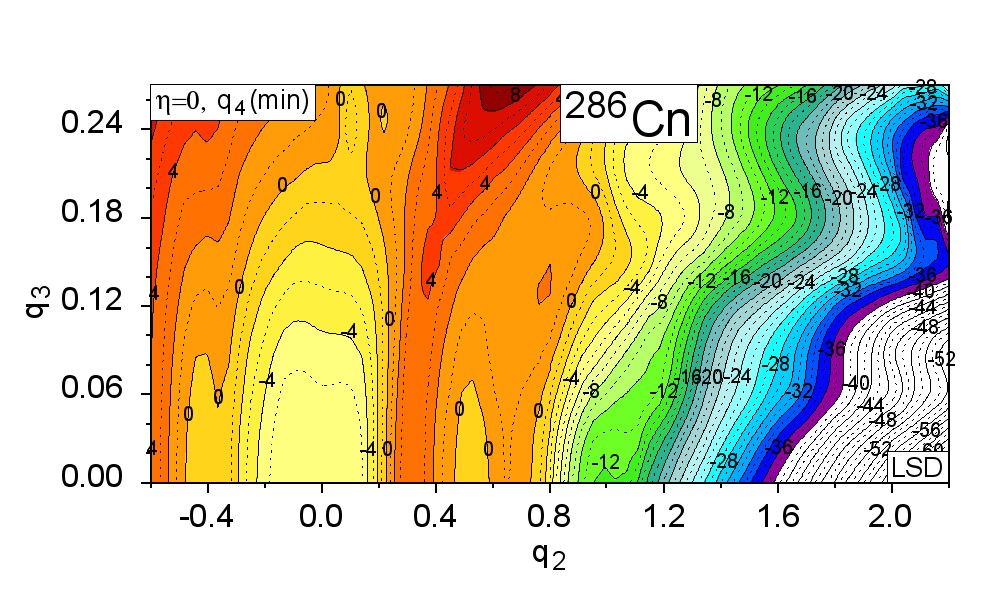}\\
\includegraphics[width=0.95\columnwidth]{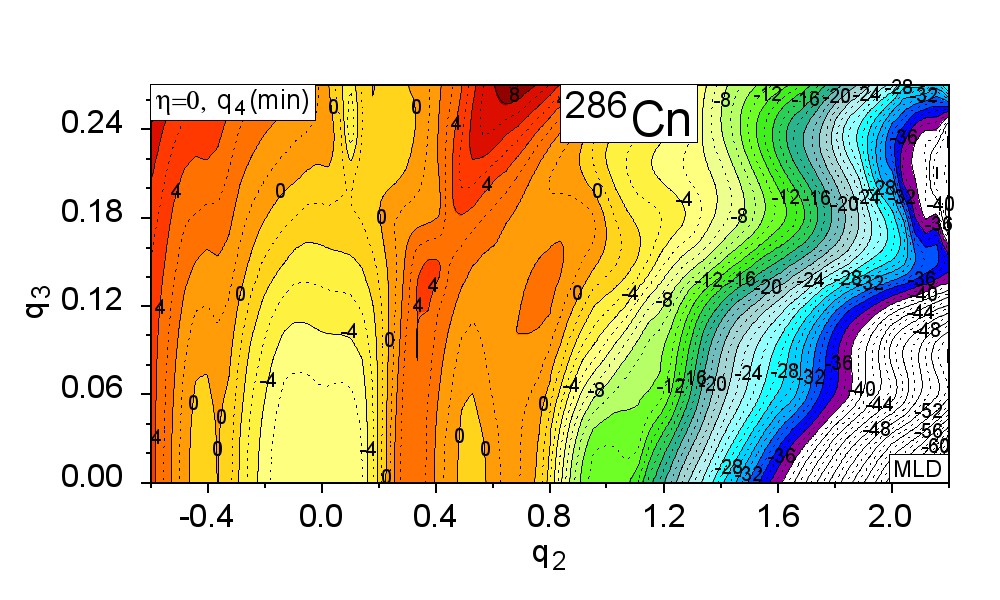}
\caption{PES obtained using the LSD (top) and MLD liquid-drop (bottom) formula of
$^{286}$Cn on the ($q_2,q_3$) plane.}
\label{286Cn23}
\end{figure}

The potential energy surface of $^{286}$Cn in the $(q_2, q_3)$ plane is shown
in Fig.~\ref{286Cn23}. The top panel corresponds again to the case where the
LSD mass formula (\ref{etot}) is used, while the bottom part is obtained with
the MLD liquid drop energy, Eq.\ (\ref{mld}). Both PES cross-sections are found
to be very similar. A slightly smaller stiffness in the $q_3$ direction is
observed in the MLD results at small elongations $q_2$, where the beginning of
a valley which could probably lead to an $\alpha$-decay appears. We speak here
about the {\it beginning} of such a valley since some more deformation
parameters would be needed to describe with reasonable accuracy such a decay
mode. Such valleys were also observed in the Gogny-HFB calculations presented in
Ref.~\cite{MZR18}. The main fission mode of $^{286}$Cn is a symmetric one since
for a given $q_2$, the minimum of the energy corresponds to a reflection 
symmetric shape ($q_3=0$). Apart from this symmetric path to fission, two
asymmetric valleys appear in Fig.~\ref{286Cn23}. One is at $q_3=0.08$, and
the second is very asymmetric ($q_3\approx 0.20$), which corresponds to a mass
of the heavy fragment around A\,=\,208. Note that the deformation-energy
estimates obtained for very elongated shapes, close to the scission
configuration ($q_2\approx 2.2$), are very close to each other in the LSD and
MLD models. 
\begin{figure}
\includegraphics[width=0.95\columnwidth]{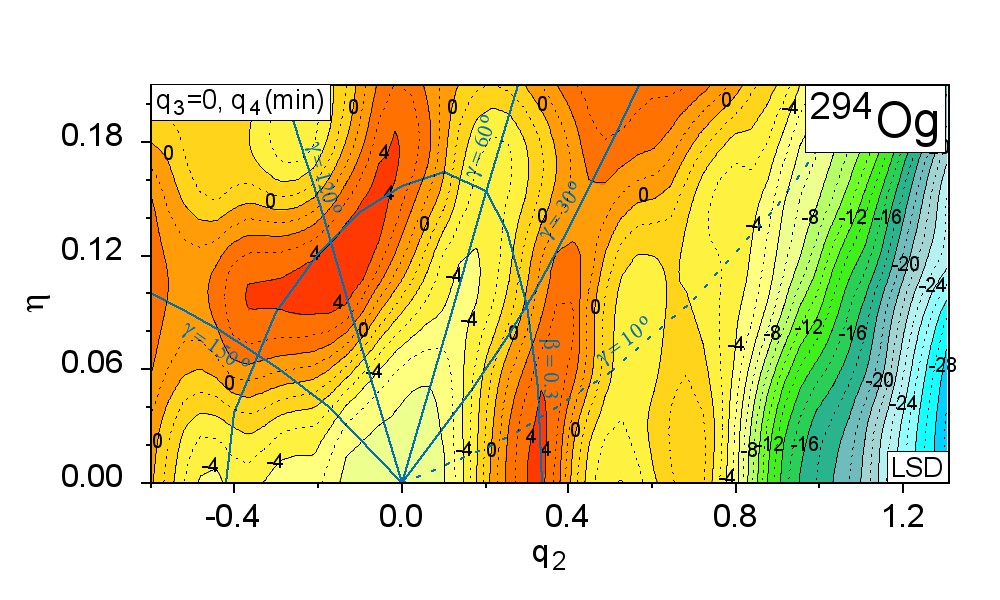}\\
\includegraphics[width=0.95\columnwidth]{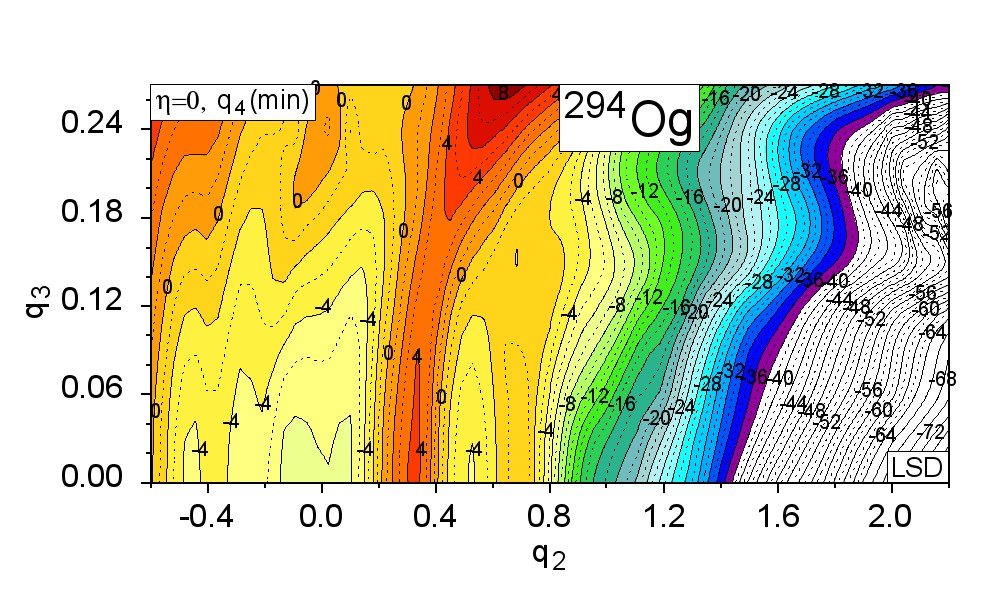}
\caption{PES obtained using the LSD liquid-drop formula of $^{294}$Og on the
($q_2,\eta$) (top) ($q_2,q_3$) (bottom) planes.}
\label{294Og23}
\end{figure}

Similar cross-sections of the PES of the heaviest synthesized nucleus 
$^{294}$Og are shown in Fig.~\ref{294Og23}. Here we present only the results 
obtained with the LSD macroscopic energy since the MLD model gives very similar
PES. A pronounced reduction of the saddle point energy, more significant than in
the case of $^{286}$Cn, due to the breaking of axial and left-right reflection 
symmetries, is predicted in this isotope. The path to fission goes from a 
spherical ground-state via oblate-shapes, a triaxial and left-right asymmetric
first saddle, a symmetric second minimum, and an asymmetric second saddle. Two
fission valleys, the deeper one symmetric, the other one very asymmetric,
corresponding to a mass of the heavy fragment around A\,=\,208, lead to the
scission configuration. 
\begin{figure}[h!]
\includegraphics[width=\columnwidth, angle=00]{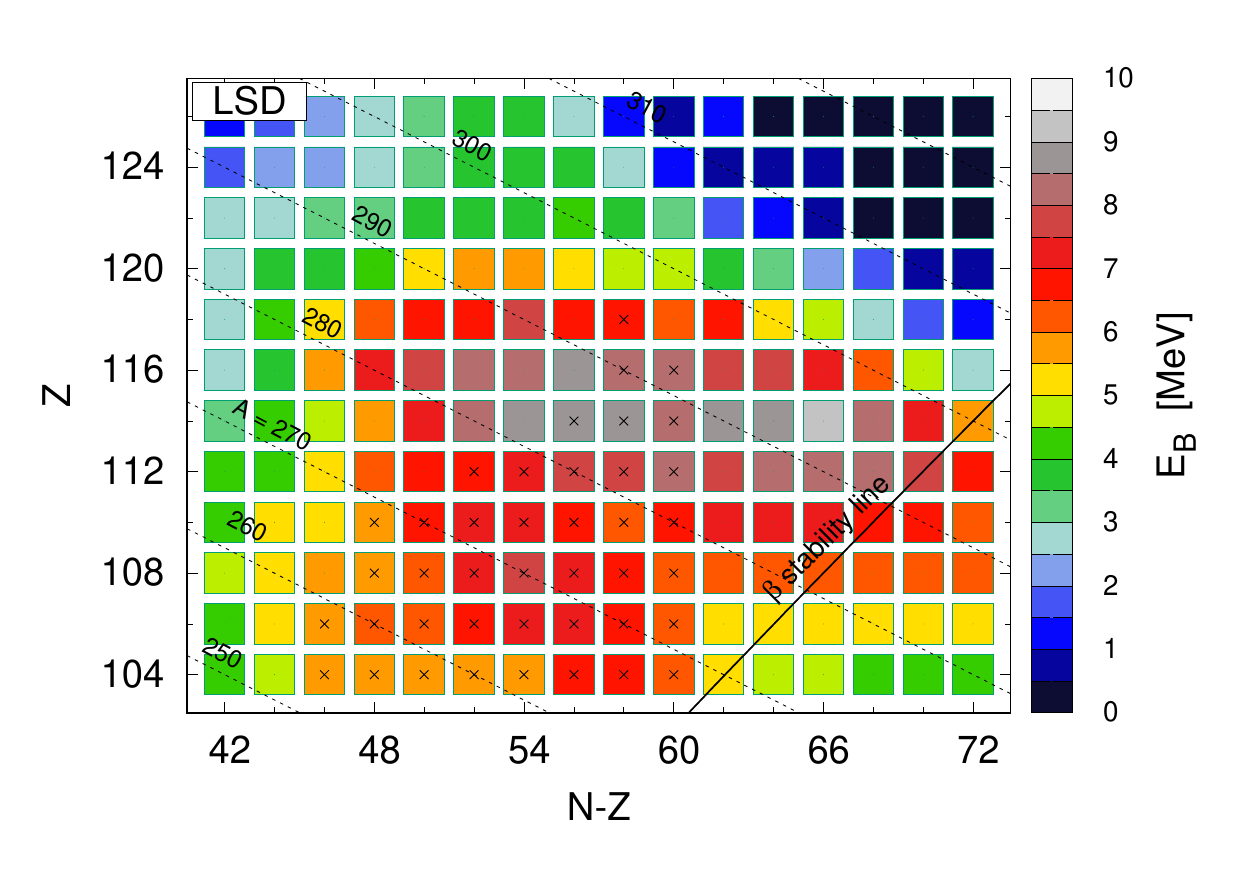}
\caption{Fission barriers heights of even-even superheavy nuclei with 
104 $\le $Z$ \le $126.}
\label{barriers}
\end{figure}

A significant reduction of the fission barrier height due to the non-axial and
reflection degrees of freedom is observed in most superheavy nuclei. In
Fig.~\ref{barriers} the fission-barrier heights $E_{\rm B}$ obtained with the
LSD model for nuclei with $104\le Z\le 128$ and $250\le {\rm A}\le 324$ are
displayed in the (A,\,Z) plane. These barrier heights defined as the difference
between the highest saddle point and the ground state energy are evaluated using
the flooding technique in the 4D $(q_2,\,\eta,\,q_3,\,q_4)$ deformation
parameter space. As one can see, the highest barriers exceeding 8 MeV are found
in the region with $112\le {\rm Z}\le 118$ and $280\le {\rm A} \le 294$. Above
${\rm A}\approx 310$ the barriers practically vanish. The experimental estimates
of the lower limit of the barrier heights obtained in Ref.~\cite{IOZ02} for a
few SHN are somewhat smaller than our predictions.
\begin{figure}[h!]
\includegraphics[width=\columnwidth, angle=00]{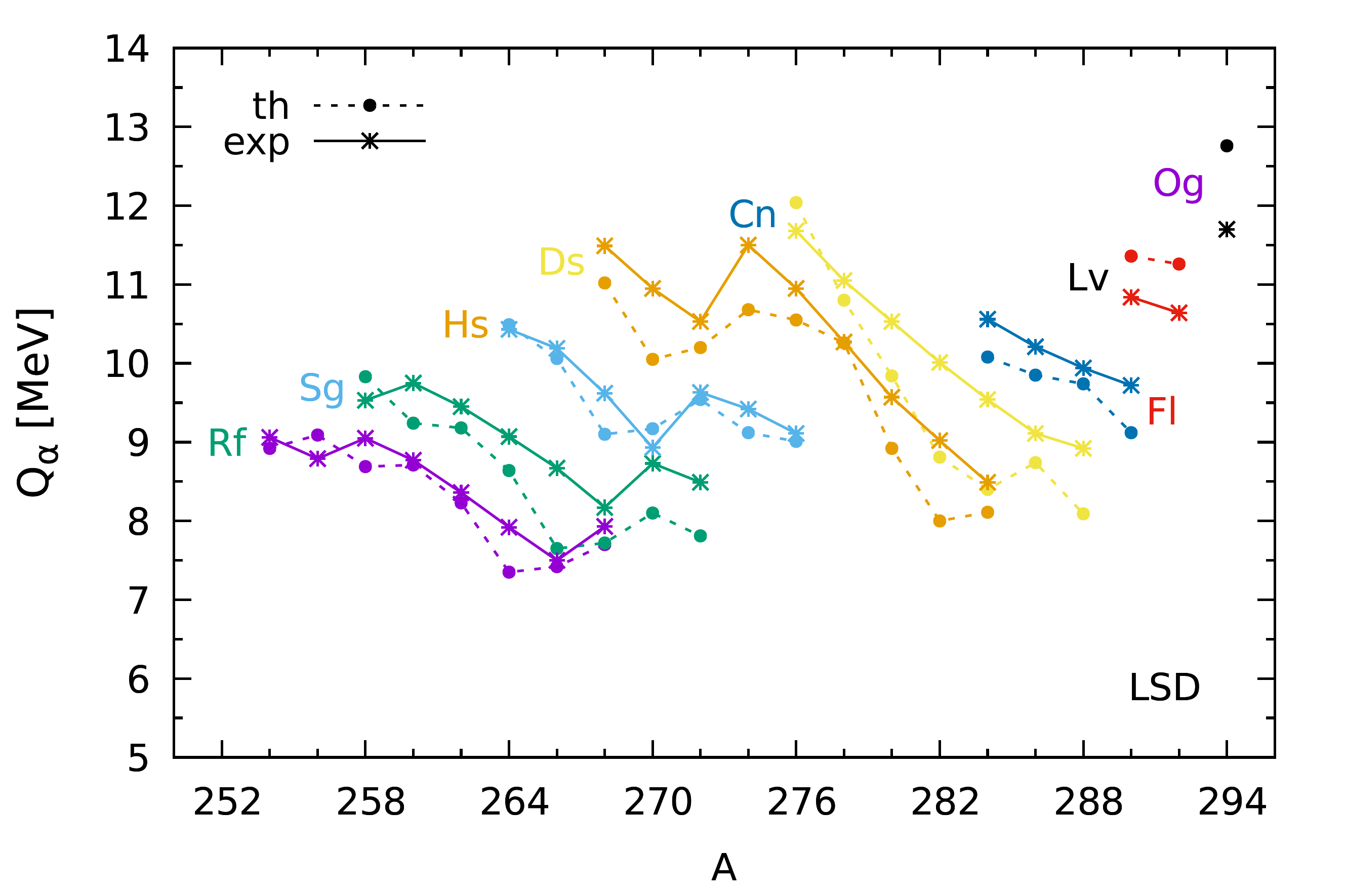}
\caption{Alpha decay energies $Q_\alpha$ of even-even superheavy nuclei with
104 $\le$ Z $\le$ 126.}
\label{qalpha}
\end{figure}

The theoretical values of the $Q_\alpha$ energies evaluated from the predicted
mass difference of nuclei $^{\rm A}$X$_{\rm Z}$ and $^{\rm A-4}$Y$_{\rm Z-2}$
are compared in Fig.~\ref{qalpha} with the experimental data (crosses) taken
from Ref.~\cite{nudat}. The agreement of the theoretical values with the data is
generally satisfactory, but with some exceptions, where the differences exceed 1
MeV.

 
\section{Spontaneous fission lifetimes in a simple model}

\begin{figure}[h!]
\includegraphics[width=0.95\columnwidth, angle=0]{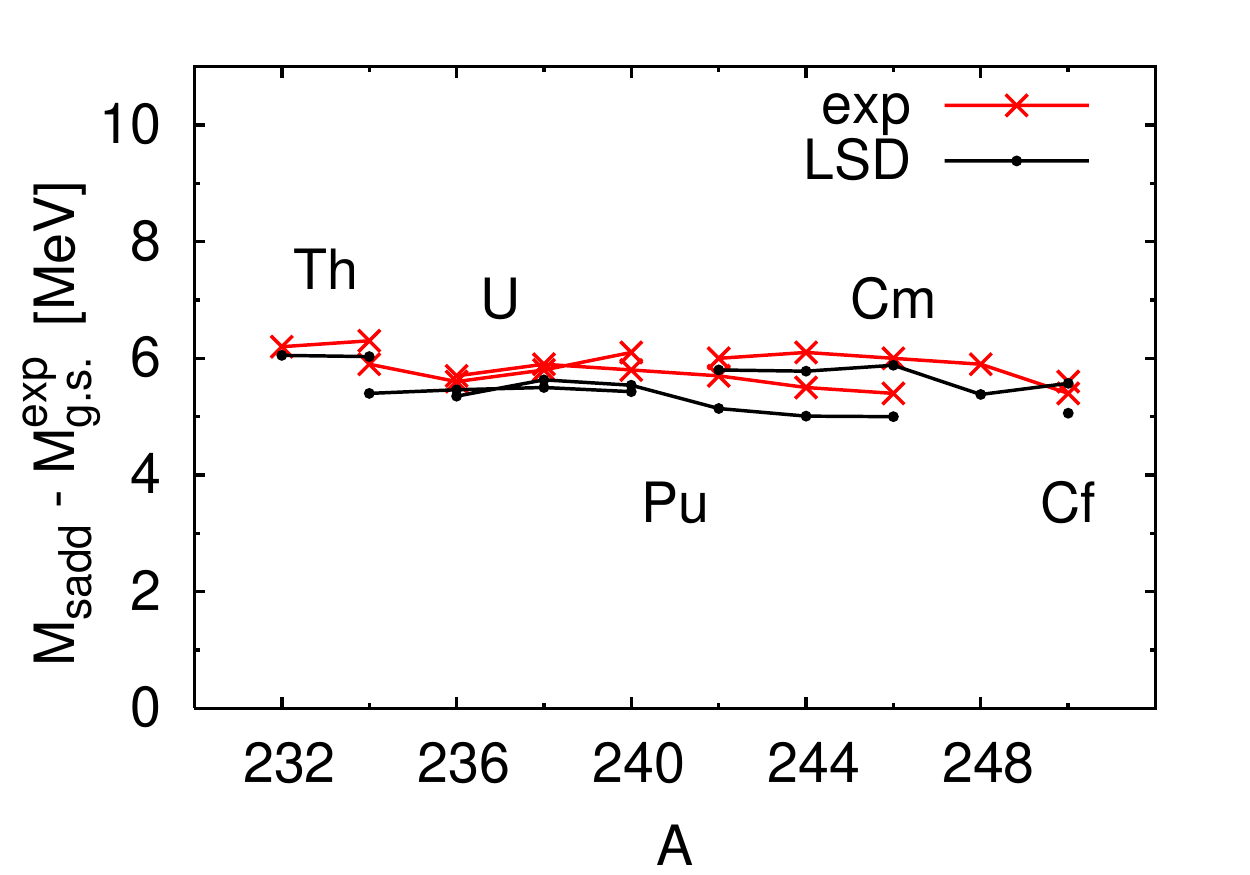}
\caption{Experimental fission barrier heights compared with the difference of
the LSD saddle and the measured nuclear mass in the ground-state \cite{BDP07}.}
\label{topo}
\end{figure}

Let us remind the reader of the so-called topographical theorem of Myers and 
\'Swi{\c a}tecki, which states that {\it the mass of a nucleus at the saddle
point is approximately equal to its macroscopic estimate}. According to this
statement, the fission barrier is equal to the difference between the saddle
mass of the nucleus obtained in the macroscopic model and its experimental mass
in the ground state. To verify that statement, a corresponding calculation was
performed in Ref.~\cite{BDP07} where the LSD mass formula, Eq.\ (\ref{lsd}), was
used to describe the macroscopic part of the binding energy. The fission
barriers extracted experimentally for even-even actinide nuclei are compared in
Fig.~\ref{topo} with the estimates obtained using the topographical theorem and
the LSD mass formula.    
The results of this investigation are striking and prove the above idea
of Myers and \'Swi{\c a}tecki. The r.m.s.\ deviation between the prediction 
of such simple estimates and the data is only about 310 keV. 
\begin{figure}[ht!]
\centerline{\includegraphics[height=0.8\columnwidth, angle=90]{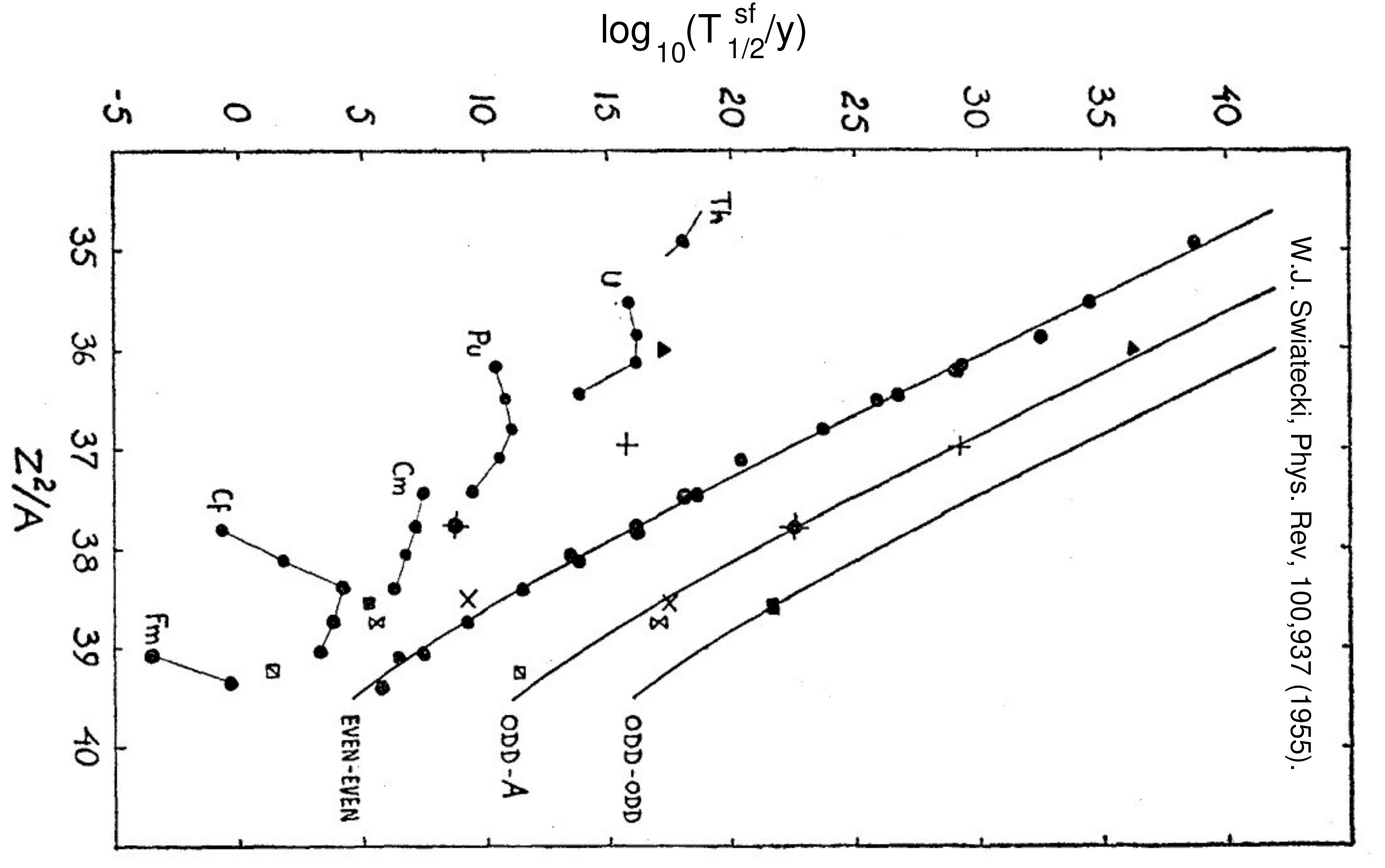}}
\caption{\'Swi{\c a}tecki systematics of the spontaneous fission lifetimes
\cite{Swi55}.}
\label{swtsf}
\end{figure}
\begin{figure}[t!]
\includegraphics[width=0.95\columnwidth]{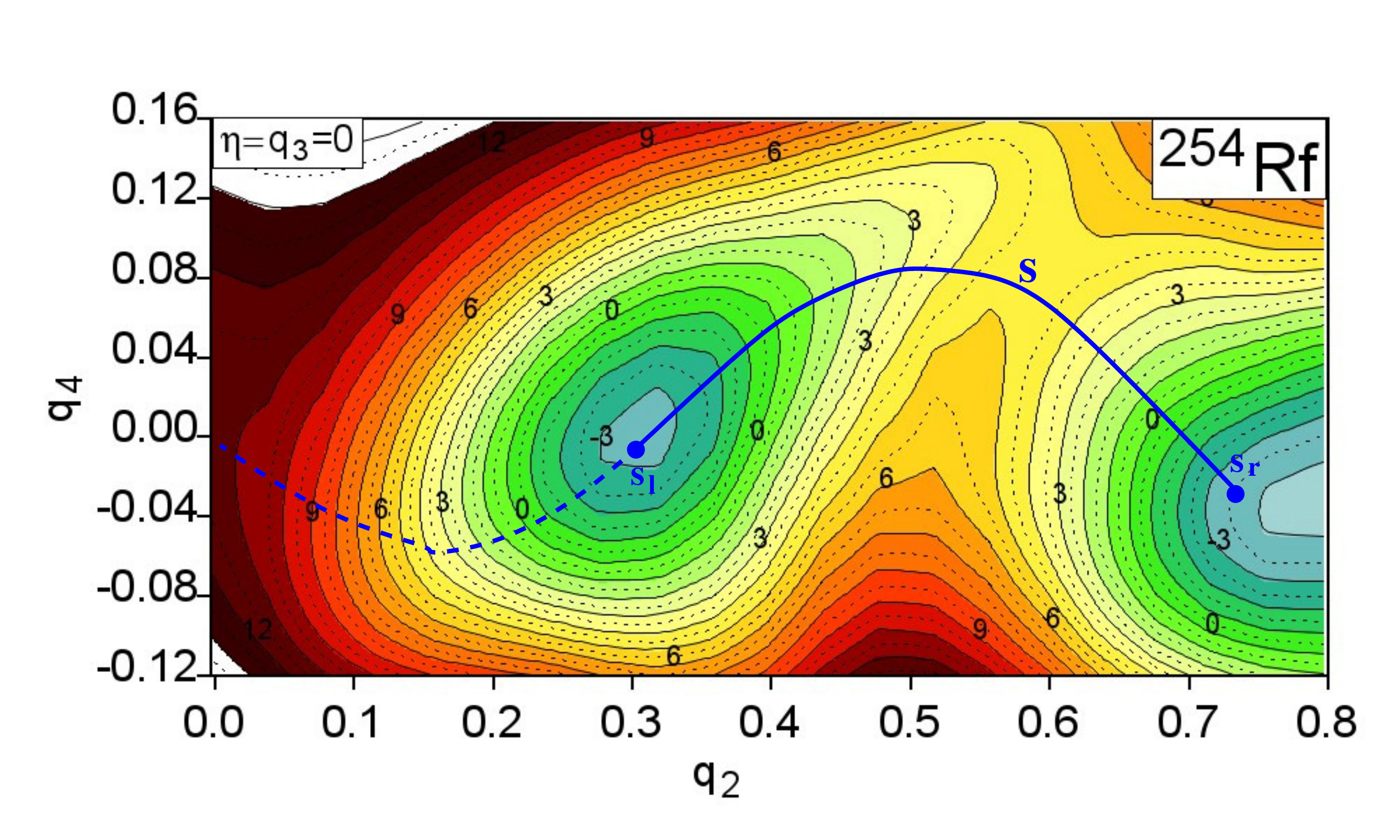}\\
\includegraphics[width=0.95\columnwidth]{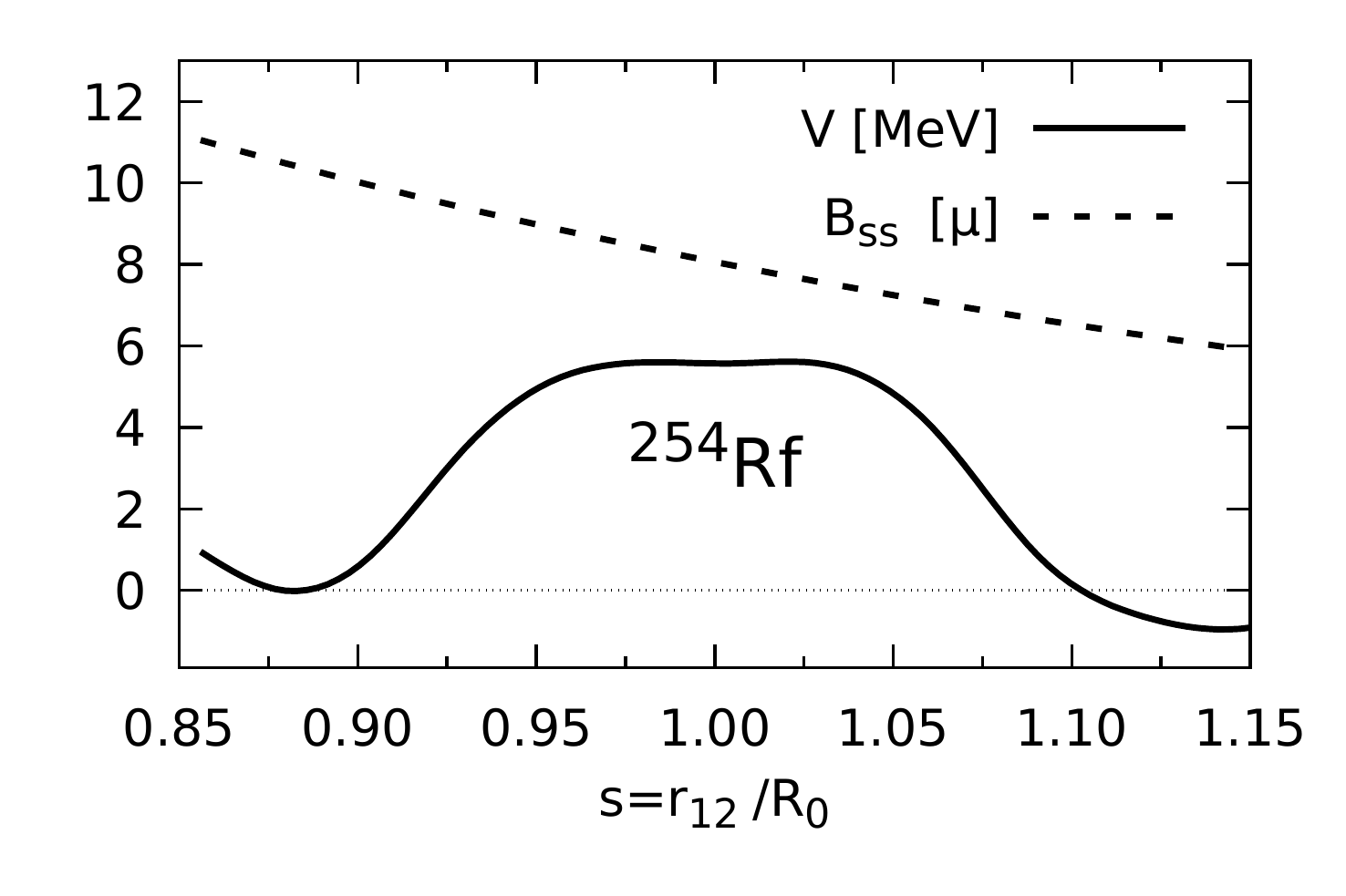}\\
\includegraphics[width=0.95\columnwidth]{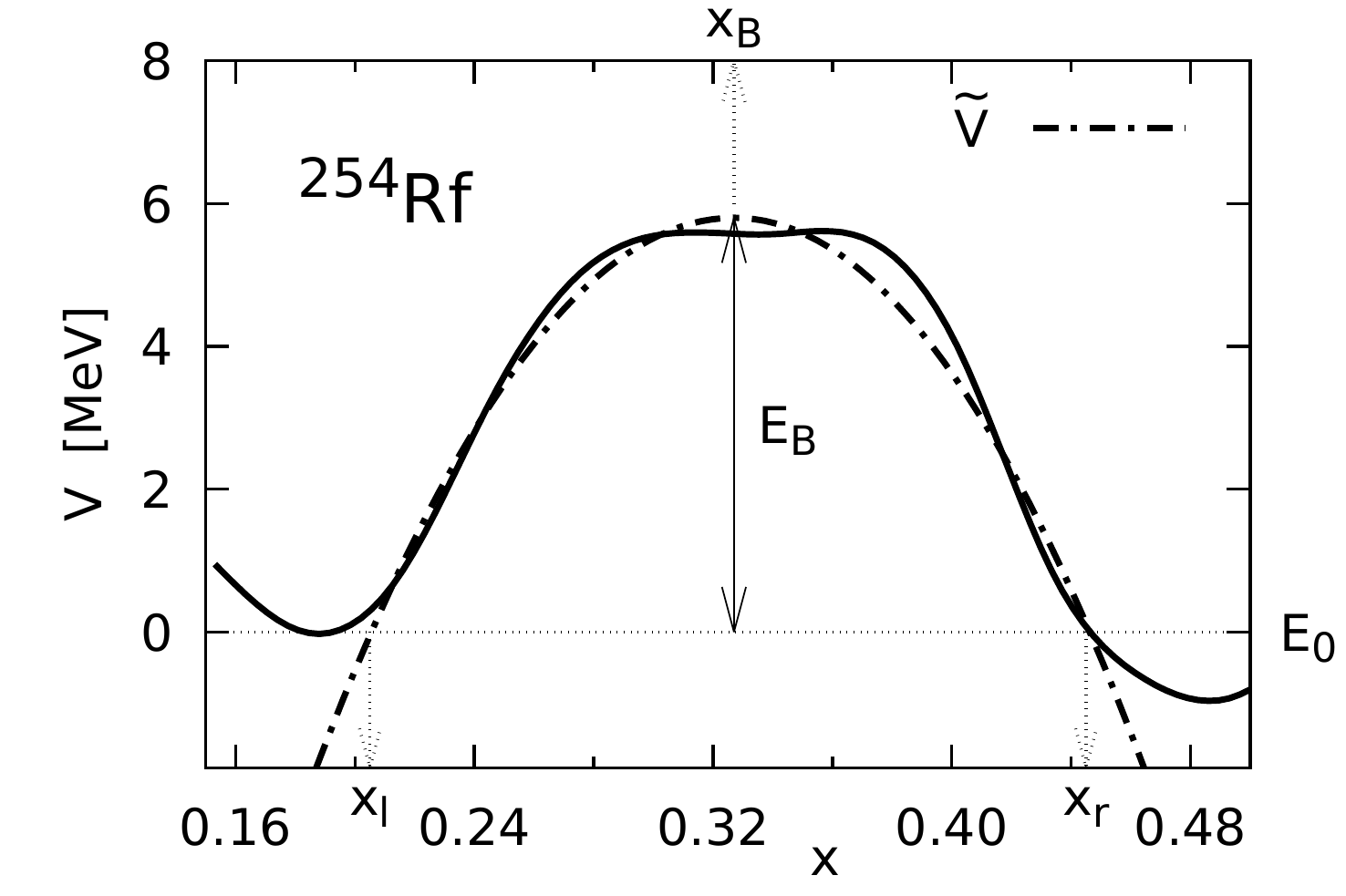}
\caption{Two-dimensional cross-section of 4D PES of $^{254}$Rf (top) with marked
static path to fission (solid blue line). The potential barrier along this path
is shown in the middle panel as a function of the relative distance between the
mass-center of the fragments $s=r_{12}/R_0$. The dashed line shows the
corresponding phenomenological mass parameter in the reduce mass units ($\mu$),
found in Ref.~\cite{RLM76}. The potential as function of the new coordinates $x$
in which the mass parameter is constant ($B_{xx}=6\mu$) is drawn in the bottom
panel. The dash-dotted line shows the approximation of this potential by two
parabolas.} 
\label{parab}
\end{figure}

In 1955 \'Swi{\c a}tecki has made a famous systematic analysis \cite{Swi55} of 
the spontaneous fission lifetimes. He has noticed, in particular, that the
quantity $\log_{10}\left(T^{\,\rm sf}_{1/2}/{\rm y}\right) + k\,\delta M$, where
$\delta M=M_{\rm exp}-M^{\rm sph}_{\rm LD}$ and an adjustable parameter $k$, is
almost a linear function of the fissility parameter for even-even nuclei. The
lines representing the results for odd-A and odd-odd nuclei are then simply 
shifted by the so-called hindrance factor, as one can see in Fig.~\ref{swtsf} 
taken from Ref.~\cite{Swi55}. It was shown in Refs.~\cite{KPo12,ZWP15} that
\'Swi{\c a}tecki like systematics of spontaneous fission half-lives work
surprisingly well for up-to-date experimental data for actinide nuclei up to
Z=102, when one uses the LSD estimates of spherical nuclei \cite{ZWP15}:
\begin{equation}
\log_{10}[T^{\,\rm sf}_{1/2}/{\rm y}]=-4.1\,Z+380.2 -7.7\,\delta M+ h~,
\label{ztsf}
\end{equation}
where $h$ is the hindrance factor equal to 0, 2.5, and 5 for even-even, odd-A,
odd-odd nuclei, respectively.

The question arises, why the \'Swi{\c a}tecki prescription for $T^{\,\rm
sf}_{1/2}$ works so well? To answer this question, let us construct a simple
model. The fission barrier of a given nucleus can be found by analyzing the
static (minimal energy) or the dynamic (minimal action integral) \cite{BPL81}
path of the PES in the multidimensional $\{q_i\}$ space, like the one presented
in the top part of Fig.~\ref{parab}, where the ($q_2,q_4$) cross-section of the 
4D PES of $^{254}$Rf is shown. The path (thick blue line) begins at the 
ground-state equilibrium point ($s_l$) and runs up to the exit point ($s_r$), 
where the energy is equal to the ground-state energy $E_0$. The fission barrier
obtained along the static trajectory, where the energy has been minimized with
respect to all deformation degrees of freedom except $q_2$, is presented in the
middle panel of Fig.~\ref{parab} as a function of the relative distance between
the mass centers of both fragments $s=r_{12}/R_0$. The dashed line shows the
phenomenological collective inertia $B_{ss}$ that enters, through the action
integral, the calculation of the spontaneous fission lifetime (see below). It is
clear that along the fission path, each deformation parameter $q_i(s)$ is a
function of the path length $s$. The classical energy ${\cal H}$ of the nucleus
is the sum of the kinetic and potential $V$ energies:
\begin{equation}
 {\cal H} = \frac{1}{2}B_{ss}(s)\dot s^2 + V(s)~,
\label{eclas}
\end{equation} 
where $B_{ss}$ is the mass parameter and $V$ the collective potential energy
along the path $s$. A simple transformation from the $s$ to a new $x$ coordinate
which conserves the kinetics energy:
\begin{equation}
x(s)=\int\limits_{s_{\rm sph}}^s \sqrt{\frac{B_{ss}(s')}{m}}\, ds'\,\,,
\label{stox}
\end{equation}
ensures that the mass parameter $B_{xx}=m$ corresponding to the new coordinate 
remains constant. One has assumed here that $x=0$ for the sphere. This
transformation to a constant mass parameter is also valid in the quantum
Hamiltonian when one does not take into account the derivatives of the
$B_{ss}(s)$ parameter. The potential $V[s(x)]$ in the new coordinate $x$, as
shown in the bottom panel of Fig.~\ref{parab}, can be approximated by
two (or more like in Ref.~\cite{RLM76}) parabolas, one for each side of the
barrier:
\begin{equation}
\widetilde V(x)= \left\{
\begin{array}{ll}
V_{\rm sadd} - \frac{1}{2}\,C_l\, (x-x_{\rm B})^2~&{\rm for}
                                           ~x\,<\,x_{\rm B}\,\,,\\[1ex]
V_{\rm sadd} - \frac{1}{2}\,C_r\, (x-x_{\rm B})^2~&{\rm for}
                                           ~x\,>\,x_{\rm B}\,\,,
\end{array}\right.
\label{Vx}
\end{equation}
how it is shown in the bottom part of Fig.~\ref{parab}.

\noindent
The spontaneous fission half-life is then given by:
\begin{equation}
T^{\,\rm sf}_{1/2}=\frac{\ln2}{nP}~,
\end{equation}
with
\begin{equation}
P=\frac{1}{1+\exp\{2S(L)\}}~,
\end{equation}
where the WKB action integral along the fission path $L(x)$ is given by:
\begin{equation}
\begin{array}{rl}
S(L) &= \int\limits^{s_r}_{s_l} \sqrt{{2 \over \hbar^2} \,
 B_{ss}[V(s) - E_0]}\, ds \\[+4ex]
&\approx
\!\int\limits_{-x_l}^{x_r}\!\!\sqrt{\frac{2m}{\hbar^2}[\widetilde
V(x)-E_0]}\,dx~.
\end{array}
\end{equation}
The penetration probability of the two (inverted parabola) barriers 
of height $E_{\rm B}$ is equal to:
\begin{equation}
\begin{array}{rl}
S=&\frac{\pi}{2\hbar}E_{\rm B}\left(\sqrt{\frac{m}{C_l}}
    +\sqrt{\frac{m}{C_r}}\right)\\[+3ex]
 &=\frac{\pi}{\hbar}E_{\rm B}\,\frac{\omega_l+\omega_r}{2\,\omega_l\,\omega_r}
 \equiv \frac{\pi}{\hbar}E_{\rm B}\,\tilde\omega^{-1}\,\,,
\end{array}
\end{equation}
where $\omega_l=\sqrt{C_l/m}$ and $\omega_r=\sqrt{C_r/m}$ are the inverted
harmonic oscillator frequencies.

\noindent
For $S\!\gg\! 1$ the logarithm of the spontaneous fission half-lives takes the
form:
\begin{equation}
\begin{array}{rl}
  \log_{10}& (T^{\,\rm sf}_{1/2})=
   \frac{2\pi}{\hbar\tilde\omega}E_{\rm B} 
  - \log_{10}[n\,{\rm ln}2]\\[+2ex]
&\approx \frac{2\pi}{\hbar\tilde\omega}(M_{\rm sadd}^{\rm LSD}-M_{\rm exp}) -
\log_{10}[n\,{\rm ln}2]\,\,,
\end{array}
\end{equation}
where $n$ is the number of assaults against the fission barrier.
The final formula for the spontaneous fission half-lives can then be written as 
\begin{equation}
\log_{10}(T^{\,\rm sf}_{1/2})+\frac{4{\delta M}}{\hbar\tilde\omega}=
\frac{4E_{\rm B}^{\rm LSD}}{\hbar\tilde\omega} - \log_{10}(n\,{\rm ln}2)~,
\label{tsfVB}
\end{equation}
where $\delta M=M_{\rm exp}-M_{\rm LSD}^{\,\rm sph}$.
The above formula can be approximated similarly as it was
done in Ref.~\cite{Swi55}:
\begin{equation}
\log_{10}(T^{\,\rm sf}_{1/2}/{\rm s})-a\,\delta M=f(E_{\rm B})~,
\label{tsfEB}
\end{equation}
where $T^{\,\rm sf}_{1/2}$ is measured in seconds ($s$), $a$ is a constant which
has to be found, and $f(E_{\rm B})$ is an adjustable function which
approximates the left side of this equation.
\begin{figure}[t!]
\includegraphics[width=\columnwidth]{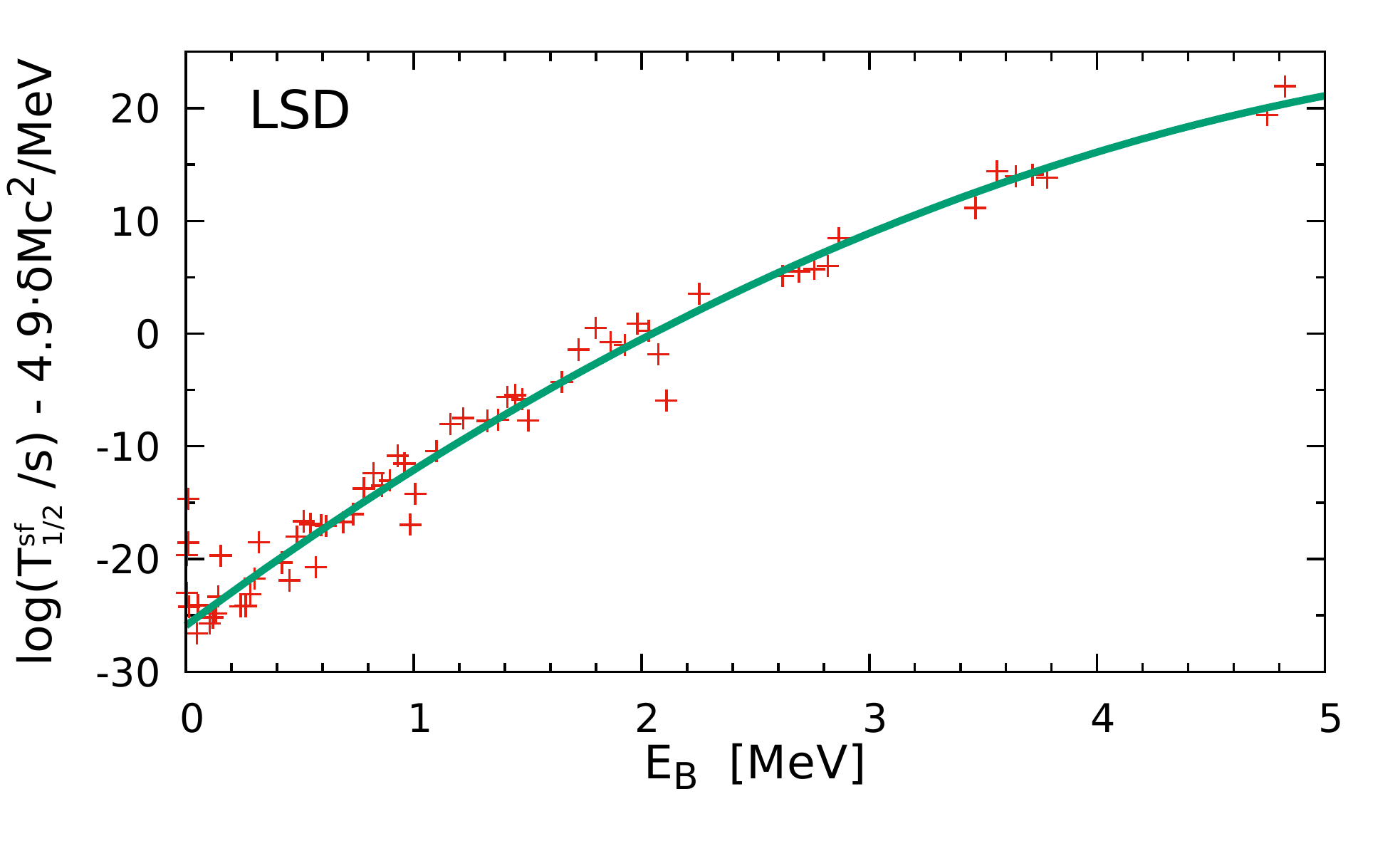}
\caption{Logarithm of the spontaneous fission half-lives corrected by the
experimental and LSD mass difference as a function of the LSD barrier height.}
\label{tsfn}
\end{figure}
\begin{figure}[h!]
\includegraphics[width=\columnwidth]{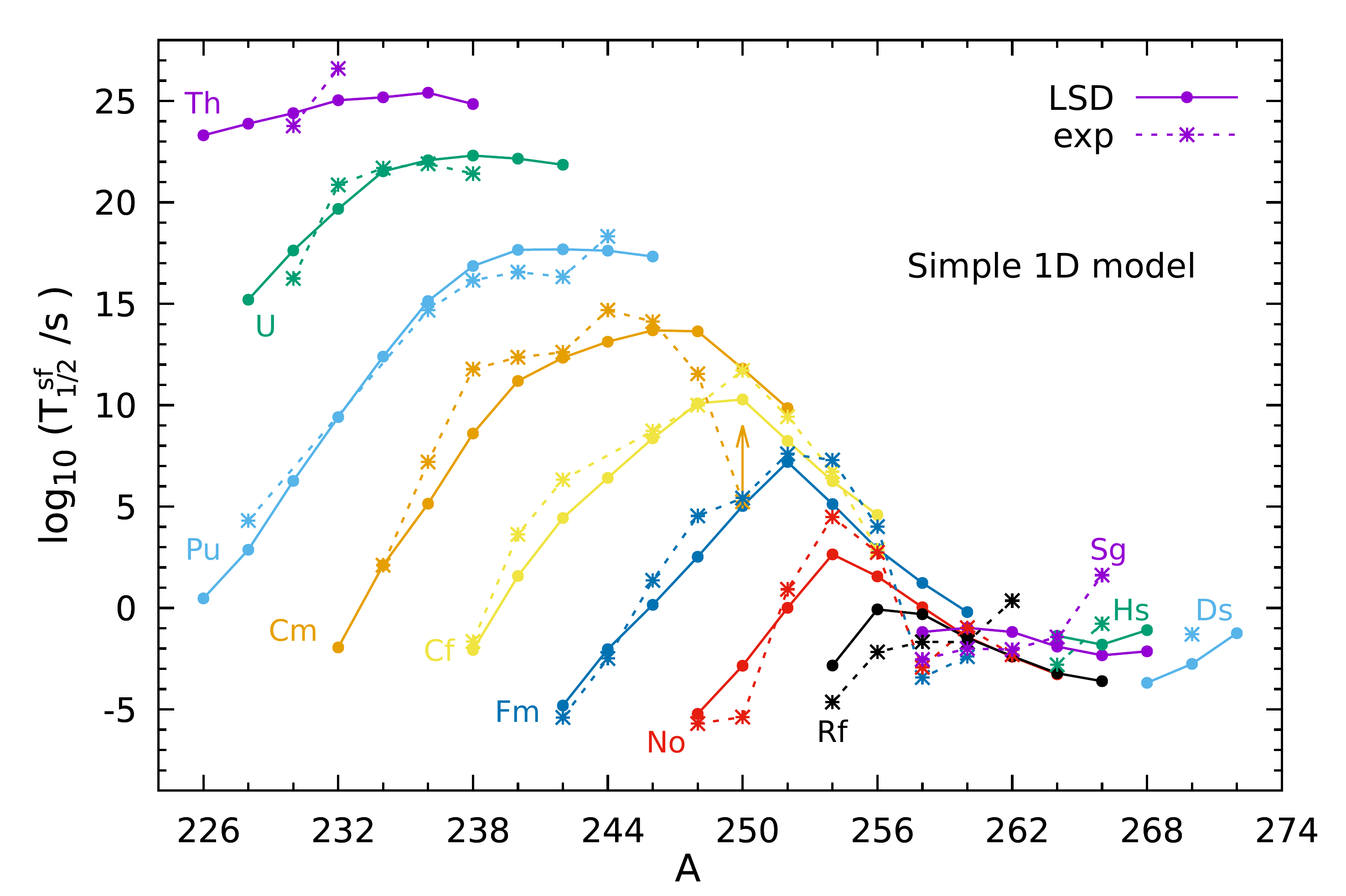}
\caption{Logarithm of the spontaneous fission half-lives predicted by formula
(\ref{final}) and the experimental data as a function of A.}
\label{tsfeb}
\end{figure}

In Fig.~\ref{tsfn} the l.h.s.\ of Eq.~(\ref{tsfEB}) is drawn as function of the
barrier height for all experimentally known $T_{\rm sf}$. The coefficient $a$ is
found equal to 4.9 MeV$^{-1}$ and the function $f(E_{\rm B})$ (solid line in
Fig.~\ref{tsfn}) is taken in the form of a second order polynomial
\begin{equation}
f(E_{\rm B})=-25.9+14.9\,E_{\rm B}-1.1\,E_{\rm B}^2~,
\label{fEB}
\end{equation}
where the coefficient of the polynomial have been obtained by a least-square
fit to the data for nuclei with Z$\,\geq\,$90. The logarithm of the spontaneous
fission half-lives can thus be approximated by
\begin{equation}
  \log_{10}(T^{\,\rm sf}_{1/2}/{\rm s})= 4.9\,(M_{\rm LSD}^{\rm sph}
  -M_{\rm exp}) + f(E_{\rm B})~.
\label{final}
\end{equation}
The above theoretical estimate of $T^{\rm sf}_{1/2}$ is compared with the 
experimental data in Fig.~\ref{tsfeb} for nuclei from Th (Z$\,=\,$90) to Ds (Z$\,=\,$110). 

As one can see, the simple one-dimensional WKB model reproduces the measured
lifetimes quite well. Similar estimates of $T_{\rm sf}$ obtained using the MLD
formula (\ref{mld}) are found to be very close to those obtained with the LSD
masses.


\section{Alpha decay lifetimes in the Gamow-like model}

The lifetimes of nuclei decaying through $\alpha$ or cluster emission can be 
estimated quite accurately using a Gamow-like model \cite{Gam28} as 
shown in Ref.~\cite{ZWP13}. Let us recall here the main assumption of this
model: The nucleus $B$ decays into two parts: $C,D$:
$$ 
^{\rm A}B_{\,\rm Z} ~\longrightarrow ~ ^{\rm A_1}C_{\,\rm Z_1}~+~
  ^{\rm A_2}D_{\,\rm Z_2}~,
$$
where the charges and the mass numbers of the daughter nucleus $C$ and of the
emitted nucleus $D$ are denoted by (Z$_1$,A$_1$) and (Z$_2$,A$_2$),
respectively.
\begin{figure}
\includegraphics[width=0.95\columnwidth, angle=0]{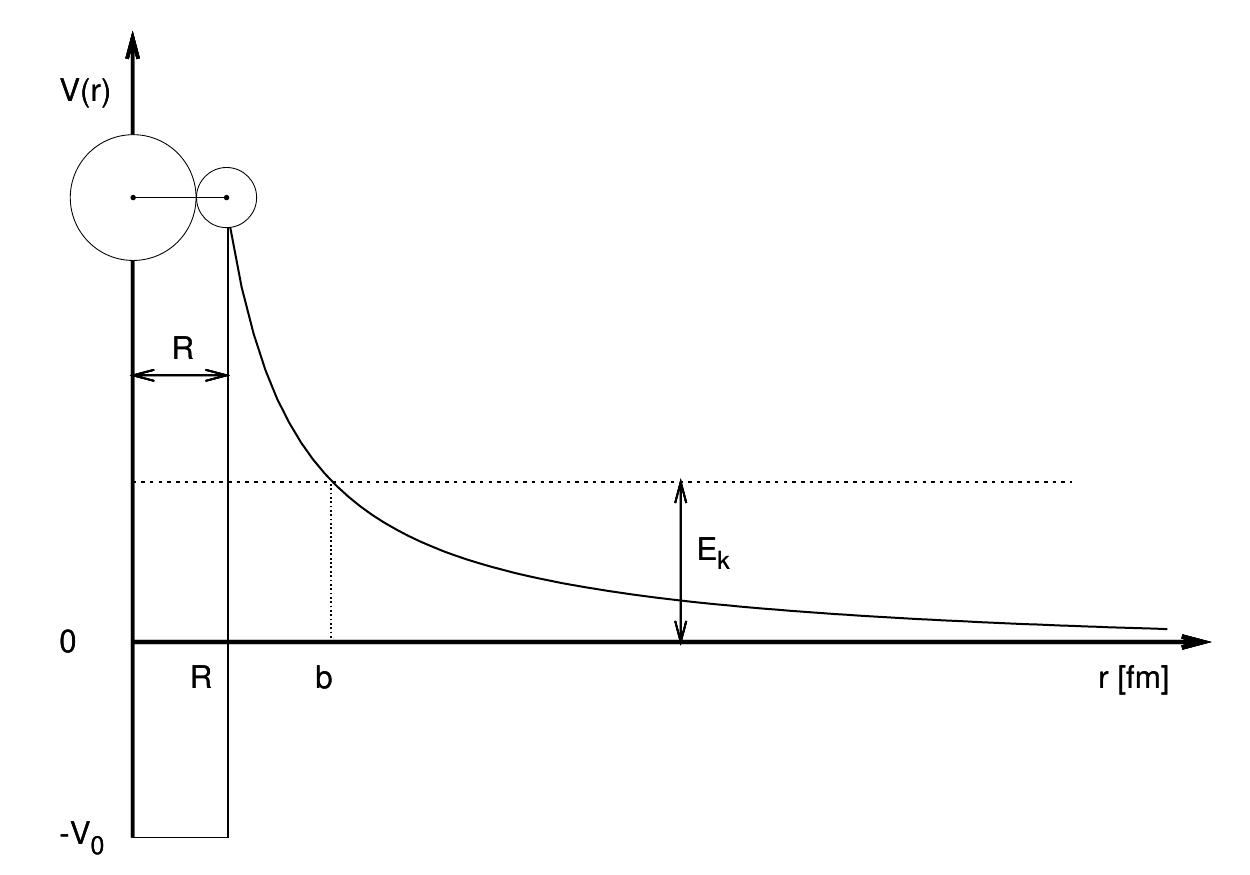}
\caption{Potential barrier tunneling by $\alpha$ particle}
\label{Valpha}
\end{figure}

A rectangular nuclear potential of depth $V_0$ and radius $R$ for the nuclear
part and a Coulomb potential $V(r)$ for the outer part, as assumed in 
Ref.~\cite{ZWP13}, is shown in Fig.~\ref{Valpha}. The emitted nucleus, as
e.g.\ an $\alpha$-particle, has the energy $E_k$. The exit point from the
barrier is denoted by $b$. The decay half-life is then given by:
\begin{equation}
T^{\,\alpha}_{1/2}=\frac{\ln 2}{\lambda} \cdot 10^h
\label{t12a}
\end{equation}
where $\lambda$ is the decay width and $h$ is a hindrance factor needed to
describe the decay of odd-even or odd-odd nuclei ($h=0$ for even-even nuclei).
The decay width can be written as  
\begin{equation}
 \lambda=\nu P~,
\label{nuP}
\end{equation}
where $\nu$ is the number of assaults against the barrier and $P$ is the 
barrier penetration probability. In the WKB approximation \cite{WKB26} this 
probability is expressed by:
\begin{equation}
P=\exp{\biggl[-\frac{2}{\hbar}\int_{R}^{b} \sqrt{2\mu(V(r)
  -E_{\rm k})}\,dr\biggr]}~,
\label{Palpha}
\end{equation}
where $\mu$ is the reduced mass of the emitted particle. The exit point from
the barrier $b$ corresponds to the place where the Coulomb potential is equal 
to the kinetic energy $E_{\rm k}$:
\begin{equation}
b= {e^2{\rm Z}_1{\rm Z}_2\over E_{\rm k}}
\label{exitp}
\end{equation}
with $e^2 \!\!=\!\!1.44\,{\rm fm\cdot MeV}$ beeing the square of the elementary
charge.
\begin{figure}[t!]
\includegraphics[width=0.95\columnwidth, angle=0]{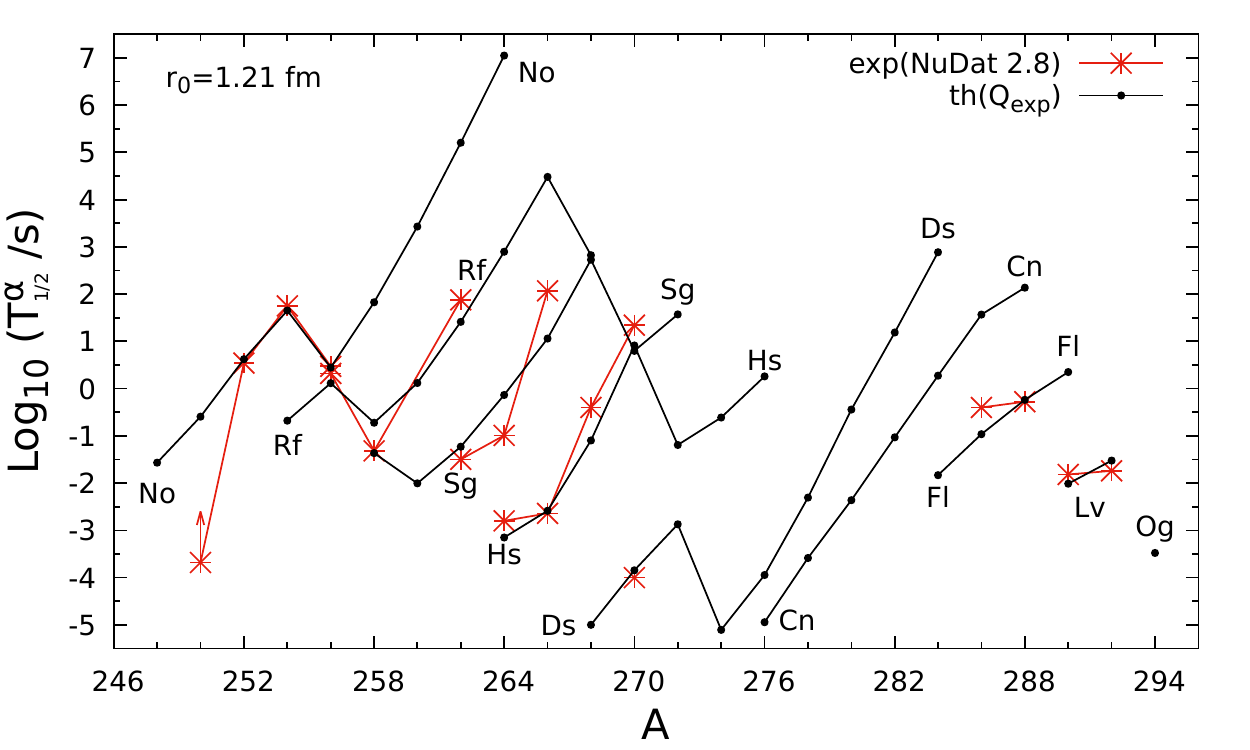}
\caption{Theoretical estimates $\alpha$ decay half-life times of even-even 
heavy nuclei compared with the data (crosses). The experimental values of
$T^\alpha_{1/2}$ and $Q_\alpha$ decay energies are taken here from 
Ref.~\cite{nudat}.}
\label{talfa}
\end{figure} 

The probability of tunneling of the Coulomb barrier of a spherical nucleus is 
given by:
\begin{equation}
\begin{array}{rl}
P\;=&\exp\left\{-\frac{2}{\hbar}\sqrt{2\mu Z_{1}Z_{2}e^{2}b}\right.\\
 \cdot& \left.\left[\arccos\sqrt{R/b}
  -\sqrt{{R/b}-\left({R/b}\right)^2}\right]\right\} \; .
\end{array}
\label{Psphere}
\end{equation}
Here $R= r_0 (A_1^{1/3}+A_2^{1/3})$ is the radius of the square well shown in
Fig.~\ref{Valpha}. It was assumed in Ref.~\cite{ZWP13} that the emitted particle
is, in its ground-state, in a square well with relatively high walls like 
presented in Fig.~\ref{Valpha}. This assumption allows to use of the number of
assaults per time-unit against the barrier corresponding to the ground state
frequency of an infinite square well:
\begin{equation}
  \nu\approx\frac{\pi\hbar}{2\mu R^2}~.
\label{nualpha}
\end{equation}
Please note that the only one adjustable parameter in this model is the radius
constant $r_0$ of the square well radius. A least-square fit performed in
Ref.~\cite{ZWP13} to all known $\alpha$-decay lifetimes of even-even nuclei has
given $r_0=1.21\;$fm. To describe the $\alpha$-decay of odd-A nuclei, an
additional hindrance factor was fitted h=0.216. For odd-odd nuclei, the
hindrance factor is simply doubled. The accuracy of reproducing the experimental
data by this simple model is merely outstanding. It turns out to be better than
that of Parkhomenko, and Sobiczewski obtained using the Viola-like formula,
which contains for even-even nuclei four adjustable parameters
\cite{ZWP13,PSo05}. A similarly good accuracy for the reproduction of the
probability of cluster \cite{ZWP13}, and proton decay \cite{ZWP16} was obtained
without any adjustment of the radius constant $r_0$. It could be mentioned that,
recently, it was shown that a careful treatment of the preformation factor of 
alpha particle in the emitter helps to improve the calculation of the 
alpha-decay width (see e.g. Ref.~\cite{XRR16}).

In Fig.~\ref{talfa} the logarithmic half-life $\log_{10}(T^\alpha_{1/2})$ for
even-even superheavy nuclei is shown, where the experimental data for the
lifetimes and $Q_\alpha$ are taken from Ref.~\cite{nudat}. It is shown that the
theoretic calculations reproduces quite well the data if available. 

Contrary to the estimates of Poenaru et al. \cite{PSG18} the cluster emission
from the SHN is orders of magnitude less probable in our model than the one for
$\alpha$-decay.

 
\section{Conclusions} 

The following conclusions can be drawn from our investigation:\\[-2ex]
\begin{itemize}
\item The Fourier expansion of nuclear shapes offers a very effective way of
      describing nuclear deformations, both at the ground-state and in the 
      vicinity of the scission configuration.\\[-2ex]
\item Two modern liquid drop models: the LSD and MLD give very close estimates 
      of nuclear masses, barrier heights, and $Q_\alpha$ energies of  
      SHN,\\[-2ex]
\item Further developments of the mean-field potentials are necessary since the
      present models predict very different magic numbers in superheavy 
      nuclei,\\[-2ex]
\item Shell and pairing effects at the ground-state determine the heights of 
      fission barriers since the influence of these microscopic effects on the 
      mass of a nucleus at the saddle point is practically negligible.\\[-2ex]
\item Spontaneous fission lifetimes of nuclei are mostly determined
      by the microscopic energy correction at the ground-state and the
      macroscopic fission barrier height. \\[-2ex]
\item A Simple WKB model with only one adjustable parameter, namely the radius
      constant $r_0$, describes well the alpha emission probabilities in SHN.
      \\[-2ex]
\end{itemize}

\noindent
Langevin type calculations, based on the mac-mic model and the 3D Fourier
shape parametrization, as well as the self-consistent method, are carried out in 
parallel by our group \cite{KDN21}. \\

\noindent
{\bf Acknowledgments}

Our research is supported by the Polish National Science Center (project No.
2018/30/Q/ST2/00185) and the National Natural Science Foundation of China
(Grants No. 11961131010 and 11790325), and the COPIN-IN2P3 agreement 
(project No.\ 08-131) between the Polish and French nuclear laboratories.

\end{document}